\documentclass[fleqn,usenatbib]{mnras}

\usepackage{newtxtext,newtxmath}

\usepackage[T1]{fontenc}

\DeclareRobustCommand{\VAN}[3]{#2}
\let\VANthebibliography\thebibliography
\def\thebibliography{\DeclareRobustCommand{\VAN}[3]{##3}\VANthebibliography}

\usepackage{graphicx}	
\usepackage{amsmath}	
\usepackage{array}
\usepackage{bm}
\usepackage[dvipsnames]{xcolor}
\usepackage{tcolorbox}
\usepackage{soul}
\usepackage[ruled,lined,linesnumbered]{algorithm2e}
\usepackage[normalem]{ulem}

\title[Cosmic shear in harmonic vs real space]{Matching Cosmic Shear Analysis in Harmonic and Real Space}

\author[Park et al.]{
Andy Park$^{1}$\thanks{E-mail: chanhyup@andrew.cmu.edu},
Sukhdeep Singh$^{1}$,
Xiangchong Li$^{1}$,
Rachel Mandelbaum$^{1}$, and Tianqing Zhang$^{1,2}$
\\
$^{1}$McWilliams Center for Cosmology and Astrophysics, Department of Physics, Carnegie Mellon University, Pittsburgh, PA 15213.\\
$^{1}$NSF AI Planning Institute, Carnegie Mellon University, Pittsburgh, PA 15213.\\
$^2$Department of Physics and Astronomy and PITT PACC, University of Pittsburgh, Pittsburgh, PA 15260, USA.
}

\date{Accepted XXX. Received YYY; in original form ZZZ}

\pubyear{2024}

\begin{document}
\label{firstpage}
\pagerange{\pageref{firstpage}--\pageref{lastpage}}
\maketitle

\begin{abstract}
Recent cosmic shear analyses have exhibited discrepancies of up to $1\sigma$ between the inferred cosmological parameters when analyzing summary statistics in real space versus harmonic space. In this paper, we demonstrate the consistent measurement and analysis of cosmic shear two-point functions in harmonic and real space using the $i${\sc Master} algorithm. This algorithm provides a unified prescription to model the survey window effects and scale cuts in both real space (due to observational systematics) and harmonic space (due to model limitations), resulting in a matching estimation of the cosmic shear power spectrum from both harmonic and real space estimators. We show that the $i$\textsc{Master} algorithm gives matching results using measurements from the HSC Y1 mock shape catalogs in both real and harmonic space, resulting in matching inferences of $S_8=\sigma_8(\Omega_m/0.3)^{0.5}$. This method provides an unbiased estimate of the cosmic shear power spectrum, and $S_8$ inference that has a correlation coefficient of 0.997 between analyses using measurements in real space and harmonic space when $S_8$ is the only free parameter. We observe the mean difference between the two inferred $S_8$ values to be 0.0004 across noise-free mock realizations, far below the observed difference of 0.042 for the published HSC Y1 analyses and well below the statistical uncertainties. While the notation employed in this paper is specific to photometric galaxy surveys, the methods are equally applicable and can be extended to spectroscopic galaxy surveys, intensity mapping, and CMB surveys.
\end{abstract}



\section{Introduction}
\label{sec:introduction}

Weak gravitational lensing (WL) is a powerful tool for investigating the large-scale structure (LSS) of the Universe. It involves the coherent deflection of light rays from distant source objects by the foreground matter distribution. Accurate WL measurements yield crucial cosmological insights into the history of structure formation, the distribution of dark matter, and the nature of dark energy. The primary advantage of this method lies in its ability to directly study the properties and distribution of dark matter through sensitivity to the matter density field along the line of sight \citep{2001PhRvD..65b3003H, 2010GReGr..42.2177H, 2013PhR...530...87W}. Cosmic shear measurements, which analyze the coherent distortion pattern in the observed shapes of distant source galaxies, stand out as an effective means to constrain structure growth and refine the cosmological model. 

The 2020s will be the most exciting decade for observational cosmology to date, due to dramatic increases in data volume as detectors, telescopes, and computers become ever more powerful. The upcoming generations of imaging surveys (known as Stage IV surveys) like the Vera C.\ Rubin Observatory Legacy Survey of Space and Time \citep[LSST;][]{2009arXiv0912.0201L, 2019ApJ...873..111I}, \textit{Euclid} \citep{2011arXiv1110.3193L}, and the \textit{Nancy Grace Roman} Space Telescope High Latitude Imaging Survey \citep{2019arXiv190205569A} will provide percent-level cosmic shear measurements by covering a wider sky coverage and collecting a larger sample of fainter source galaxies compared to previous surveys such as the Subaru Hyper Suprime-Cam \cite[HSC;][]{2018PASJ...70S...4A} survey, the Dark Energy Survey \citep[DES;][]{2016MNRAS.460.1270D}, and the Kilo-Degree Survey \citep[KiDS;][]{2017A&A...604A.134D}, and thus require us to revisit many of the assumptions made in previous studies.

Once the data is acquired and processed into shear maps or shear catalogs, the next step is to compute the two-point summary statistics. One question that we need to address at this stage is whether to carry out the cosmic shear analysis in real space or harmonic space; each space has its own advantages and disadvantages, and we introduce the trade-offs in Sec.~\ref{sec:bkgd:cosmic_shear_analysis}. Analysis in real space corresponds to measuring the two-point correlation functions of the galaxy shear estimates as a function of angular separation \citep[e.g., as in][]{2023arXiv230400702L, 2022PhRvD.105b3515S, 2022A&A...664A..77S}, and analysis in harmonic space corresponds to measuring the cosmic shear power spectrum as a function of multipole moments \citep[e.g., as in][]{2023arXiv230400701D, 2022MNRAS.515.1942D, 2022A&A...665A..56L}. One major challenge is to account for the survey window. The survey window is a function that captures the effects of the survey geometry and other observational effects that affect the selection of galaxies included in our samples. The survey window complicates the power spectrum analysis as the survey mask couples Fourier modes, which would otherwise be independent. This leads to mode coupling in the power spectrum, which is commonly addressed using pseudo-$C_\ell$ estimators \citep[e.g.,][]{2001PhRvD..64h3003W, 2011MNRAS.412...65H, 2019MNRAS.484.4127A}.

Since the coherent distortion of galaxy shapes is caused by a scalar gravitational potential of the matter distribution, we can further separate the lensing signal into curl-free ($E$ mode) and divergence-free ($B$ mode) components. The survey window also introduces leakage of $E$ modes into the $B$ mode component of a given map, and vice versa, an effect that needs to be modeled and corrected using detailed knowledge of the window. The modeling of the survey window in harmonic space was first addressed in \cite{2002ApJ...567....2H}, who introduced the \textsc{Master} algorithm to deconvolve the effects of the survey window to reconstruct the power spectrum of the underlying fields for Cosmological Background Microwave (CMB) analysis \citep{2001PhRvD..64h3003W, 2005MNRAS.360.1262B, 2001PhRvD..65b3505L}. \cite{2019MNRAS.484.4127A} introduced the now standard \textsc{NaMaster} software with an implementation of the \textsc{Master} algorithm for spin-0 and spin-2 fields that is also used for galaxy clustering and cosmic shear analyses.

Throughout this paper, we define two estimators as \textit{matching} if they yield results that agree within a percent when applied to the same dataset, and use the term \textit{biased} to describe cases where systematic effects cause a significant deviation from the expected result. For example, \cite{2020PASJ...72...16H} and \cite{2021MNRAS.503.3796D} show how such a disparity in approach produces an apparent mismatch between the analyses in real space and harmonic space when performing cosmological parameter inference. Crucially, if we truly believe certain scales cannot be well modeled, they should be removed regardless of the space in which the analysis is performed.

The survey window also affects the two-point correlation functions in real space, although this effect is typically ignored since the commonly-used estimator removes most of the window effects (see appendix~D of \cite{2020MNRAS.491...51S} for demonstration of residual percent-level window effects on correlation functions).

After measuring the summary statistics, we apply scale cuts on the measurements to restrict our analysis to scales where the model is reliable (normally done in harmonic space where the models are defined) and to remove scales contaminated by observational systematics (normally done in real space to account for physical and observational effects tied to angular separation) before performing cosmological parameter inference. 

We can typically model a limited range of scales that is smaller than the range of scales probed by the current data. Thus, we need to apply scale cuts carefully to extract as much information as possible while avoiding biases from applying the model outside its range of validity. Failure to account for the scale cut that was implemented in one space when modeling the other space can lead to using different sets of information in real versus harmonic space analysis, even though the underlying field is the same. 

There have been many efforts to introduce a statistic that can separate out the $E/B$ modes. \cite{2002A&A...389..729S} introduced the band power spectra that used the aperture mass dispersion $\langle M^2_\text{ap}(\theta)\rangle$ and $\langle M^2_\times(\theta)\rangle$ \citep{1998MNRAS.296..873S} measured from a linear combination of the two-point correlation functions weighted by specific window functions to perform the $E/B$ mode decomposition. However, calculating the aperture mass dispersion requires estimation of the shear correlation functions down to zero angular separation, which is not possible in practice. To remedy this, they applied apodization in the window function that truncates the integral over all angular distances and reduces the ringing effect caused by the limited range of the two-point correlation functions. 

\cite{2010A&A...520A.116S} introduced the Complete Orthogonal Sets of E- / B-Mode Integrals (COSEBI) method to measure pure $E$ and $B$-mode cosmic shear power spectra from the two-point correlation functions over a finite interval $\theta_\text{min} \leq \theta \leq \theta_\text{max}$. The method, similar to \cite{2002A&A...389..729S}, uses a linear combination of binned measured two-point correlation functions weighted by specific window functions in the angular scale. \cite{Matth2016} introduced a similar band-power estimator for cosmic shear power spectra from linear combinations of binned two-point correlation functions. Although the aforementioned methods return unbiased two-point statistics estimators, there were still inconsistencies in analysis between the two spaces.

\cite{2021A&A...645A.104A} used three different two-point statistics (band power spectra, COSEBIs, and two-point correlation functions) on the KiDS data set and compared the measured values of the parameter $S_8 \equiv \sigma_8 (\Omega_m/0.3)^{0.5}$, where $\sigma_8$ is the standard deviation of matter density fluctuations in spheres of radius $8~\text{Mpc}/h$ (here $h \equiv H_0/100)$ and $\Omega_m$ is the fraction of matter density in the Universe with respect to the critical density. Comparing the inferred $S_8$ from each method provided a useful internal consistency check between the methods; \cite{2021A&A...645A.104A} found a $0.4\sigma$ difference in $S_8$ between the COSEBIs and the two-point correlation functions analyses. \cite{Camacho2022} did a cosmic shear analysis in real space and harmonic space using the DES-Y1 \texttt{METACALIBRATION} shear catalog, and also found $0.4\sigma$ difference in $S_8$. These past results show that offsets at the $\sim$0.5$\sigma$ level can arise because the different two-point statistics have different sensitivities to $\ell$ scales, even for the same data set.

\cite{2021MNRAS.508.1632S} further improved the \textsc{Master} algorithm (to produce $i$\textsc{Master}) and generalized the method to account for the scale cuts used in the analysis, after which the power spectra and two-point correlation function estimates  match. In this work, we present the first application of the $i$\textsc{Master} algorithm to survey mock catalogs with complex geometry. 

We measure the cosmic shear two-point functions in harmonic and real space in the HSC Y1 mock shape catalogs \citep{Shirasaki_2019}, using the $i$\textsc{Master} method to estimate the power spectrum from both harmonic and real space estimators consistently, accounting for both the survey window and the scale cuts used in the analysis in a way that ensures matching results. We quantify the consistency between the two analyses by measuring $\delta S_8 = S_{8|R\to F} - S_{8|F}$, where $S_{8|R\to F}$ is the inferred $S_8$ value going from the real space estimator to the harmonic space estimator, and $S_{8|F}$ using the harmonic space estimator directly.

This paper is organized as follows: In Sec.~\ref{sec:bkgd}, we provide an overview of cosmic shear analysis. In Sec.~\ref{sec:method}, we describe how two-point estimators are measured in harmonic and real space and outline steps to make a matching two-point estimator that includes the impact of the survey window and scale cuts. Then, in Sec.~\ref{sec:dai}, we describe the shear catalog used to test the proposed formalism in this paper and the implementation details of our estimators. The results of the reconstruction from harmonic space and real space with likelihood analysis are shown in Sec.~\ref{sec:result}. Finally, in Sec.~\ref{sec:conclusion}, we summarize our results and the future outlook.

\section{Background}
\label{sec:bkgd}
In this section, we provide a brief background to weak gravitational lensing. We introduce the history of cosmic shear in Sec.~\ref{sec:bkgd:cosmic_shear}, discuss how shear is estimated from the shapes of galaxies in Sec.~\ref{sec:bkgd:cosmic_shear_shape}, and describe how shear is used in cosmological analysis in Sec.~\ref{sec:bkgd:cosmic_shear_analysis}.

\subsection{Cosmic Shear}
\label{sec:bkgd:cosmic_shear}
Cosmic shear is the coherent distortion of images of distant background galaxies by the foreground Large Scale Structure (LSS) in the Universe (see \citealt{2015RPPh...78h6901K} for a review). Such distortions are induced by all matter along the line of sight, making cosmic shear a powerful tool for probing the matter distribution, providing us with a unique glimpse into the structure and evolution of the Universe on large scales. The first cosmic shear detections were made in the early 2000s \citep[e.g.][]{2000astro.ph..3338K, 2000MNRAS.318..625B, 2000Natur.405..143W, 2001A&A...374..757V, 2004ApJ...605...29R}, and further advancements were made in the subsequent years until now. These improvements included larger volumes of survey data, improved redshift estimation \citep[e.g.,][]{2021MNRAS.505.4249M, 2023MNRAS.524.5109R}, and refined statistical analysis \citep[e.g.,][]{2010A&A...516A..63S}. Recent surveys, as well as ongoing ones, have successfully measured cosmic shear and constrained cosmological parameters to high precision. The upcoming imaging surveys like LSST \citep{2019ApJ...873..111I}, \textit{Euclid} \citep{2011arXiv1110.3193L}, and \textit{Nancy Grace Roman} Space Telescope \citep{2019arXiv190205569A} will allow us to make percent-level measurements with unprecedented data volume. As these datasets become increasingly precise, ensuring \textit{matching} results between real-space and harmonic-space analyses is essential to prevent systematic biases in cosmological constraints.

\subsection{Shear Estimation}
\label{sec:bkgd:cosmic_shear_shape}
The distortion of galaxy shapes by weak gravitational lensing can be well described by a local coordinate transformation. The mapping between the unlensed coordinates ($x_u, y_u$) and the lensed coordinates ($x_l, y_l$) is defined according to the distortion matrix:
\begin{equation}
    \label{eq:bkgd:jacobian}
    \begin{pmatrix}
        x_u \\ y_u
    \end{pmatrix} =
    \begin{pmatrix}
        1 - \gamma_1 - \kappa & -\gamma_2 \\
        -\gamma_2 & 1 + \gamma_1 - \kappa
    \end{pmatrix}
    \begin{pmatrix}
        x_l \\ y_l
    \end{pmatrix},
\end{equation}
where $\gamma_1$ and $\gamma_2$ are the two lensing shear components and $\kappa$ is the lensing convergence. Positive (negative) $\gamma_1$ stretches the image along the $x$ ($y$) axis direction, and positive (negative) $\gamma_2$ stretches the image in the direction at an angle of $45$ deg ($-45$ deg) with the $x$ axis. The convergence $\kappa$ changes the size and brightness of the unlensed galaxy. Shear negates when rotated by $\pi/2$, making it a spin-2 quantity, whereas convergence is a spin-0 scalar quantity as it is invariant under rotation. We encode the two components of shear into a shear spinor $\gamma = \gamma_1 + i\gamma_2 = \left | \gamma\right|e^{2i\phi}$, where $i$ is the imaginary number unit, $\phi$ is the azimuthal angle, and the factor of $2$ arises due to the spin-2 nature of the shear. It is common to factor out $(1-\kappa)$ from the distortion matrix in equation~\eqref{eq:bkgd:jacobian} as this multiplier only affects the size but not the galaxy's shape. Since cosmic shear is based on the measurement of galaxy shapes and not the size, the observable that we can access in practice is not the shear $\gamma$, but rather the reduced shear $g$, which can be written in terms of $\gamma$ as
\begin{equation}
    \label{eq:bkgd:g}
    g = \frac{\gamma}{1-\kappa}.
\end{equation}
The reduced shear $g$ has the same spin-2 properties as the shear (i.e., $g = g_1 + i g_2 = |g|e^{2i\phi}$). In the weak lensing limit, where the effect of lensing is on the order of a few percent or less ($\kappa, \gamma \ll 0.02$), $g \simeq \gamma$ is a reasonable approximation to linear order \citep{2009ApJ...696..775S}. For the rest of the work, we use the reduced shear $g$ and shear $\gamma$ interchangeably. 

In cosmic shear, we use galaxy ellipticity to quantify the spin-2 aspect of galaxy shape and measure weak lensing shear distortion. In practice, we define an ellipticity spinor $e = e_1 + i e_2$ where the two components of the spinor follow the same definition as the two shear components. There exist many methods in the literature for using the light profile of a galaxy to define the ellipticity, e.g., using moments or derivatives of the galaxy's light intensity profile in harmonic space \citep{2008MNRAS.383..113Z, 2014MNRAS.438.1880B}, decompose light intensity profile onto basis functions \citep{2003MNRAS.338...48R, 2005MNRAS.363..197M, 2018MNRAS.481.4445L}, or parametric galaxy's light intensity profile model \citep{2013MNRAS.434.1604Z, 2017MNRAS.467.1627F}. We then obtain the linear shear responses of the ellipticities \citep{metacal_Huff2017, FPFS_Li2023} to infer the shear from an ensemble of galaxies with ellipticity measurements, and there are many ways to correct noise bias, selection bias and detection bias \citep{metacal_Sheldon2017, metaDet_Sheldon2020, FPFS_Li2022, FPFS_Li2023}. 

\subsection{Cosmic Shear Analysis}
\label{sec:bkgd:cosmic_shear_analysis}
The outcome of the measurements described in Sec.~\ref{sec:bkgd:cosmic_shear_shape} is a shear catalog. The shear catalog is then compressed using a summary statistic. Two common summary statistics are to measure the shear-shear power spectrum $C_\ell$ in harmonic space and the two-point correlation functions (TPCFs) $\xi_\pm(\theta)$ in real space. Higher-order statistics can also be used to extract the non-Gaussian nature of the cosmic density field not encoded by the two-point summary statistics \citep[e.g.,][]{2003ApJ...584..559Z}. Both spaces have their own advantages and disadvantages. It is often easier to estimate the TPCFs \citep{1993ApJ...412...64L} as our data is in real space, and we have a better understanding of observational systematics in real space. In contrast, Gaussian and non-Gaussian modes and linear and non-linear scales are clearly separated in harmonic space \citep{2008PhRvD..77j3013H, 2019MNRAS.484.4127A}. In harmonic space analysis, individual modes are significantly less correlated compared to real space analysis, making the covariance matrix estimation easier. Our theoretical model of the summary statistics is also computed in harmonic space. For this reason, there has been a large body of literature aimed at developing efficient power spectrum estimators, which have been widely employed in the Cosmic Microwave Background (CMB) two-point measurements \citep[e.g.,][]{2001PhRvD..64h3003W, 2005MNRAS.360.1262B, 2001PhRvD..65b3505L}. While both approaches should ideally yield \textit{matching} cosmological constraints, previous studies have found discrepancies between real space and harmonic space analyses when applying different scale cuts or improperly handling the survey function \citep{2020PASJ...72...16H, 2021MNRAS.503.3796D}. These mismatches highlight the importance of developing methods that correctly account for scale-dependent effects to avoid introducing systematic biases into cosmological inference.

The shear field is a spin-2 field, and we can further decompose the lensing signal into the curl-free ($E$ mode) and divergence-free ($B$ mode) components. We use the Limber approximation \citep{1953ApJ...117..134L} to compute the model power spectra, which is valid for scales used in this analysis \citep{2023arXiv230400701D}. 

Following this assumption, the cosmic shear power spectra for flat spatial geometry can be related to the matter power spectrum as:
\begin{equation}
    \label{eq:bkgd:ps}
    C^{EE}_{ij}(\ell) = \int_0^{\chi_H} d\chi \frac{q^i(\chi)q^j(\chi)}{\chi^2}P_m\left(k=\frac{\ell+1/2}{\chi}, z(\chi)\right),
\end{equation}
where $i$ and $j$ are tomographic bins, $\chi$ is the comoving distance, $\chi_H$ is the comoving horizon distance, and $P_m$ is the nonlinear matter power spectrum. The lensing efficiency, $q^i(\chi)$ is defined as:
\begin{equation}
    q^i(\chi) = \frac{3}{2}\Omega_m\left(\frac{H_0}{c}\right)^2\frac{\chi}{a(\chi)}\int_\chi^{\chi_H} d\chi' n^i(\chi')\frac{\chi-\chi'}{\chi'},
\end{equation}
where $H_0$ is the Hubble constant today, $a$ is the scale factor, $\Omega_m$ is the matter density parameter, and $n^i(\chi)$ is the redshift distribution of source galaxies in the $i$th tomographic bin. Since the lensing is caused by the scalar gravitational potential of the lens, the $B$ mode is zero, i.e. $C^{BB}_{ij}(\ell)=0$. However, $E$ mode leakage and some systematics can introduce $B$ modes \citep[e.g.,][]{2019MNRAS.484.4127A}. 

The cosmic shear TPCFs are expressed as a transformation of $E$ and $B$ modes of the power spectrum. With the flat sky approximation, the transformation is the Hankel transform:

\begin{equation}
    \xi^{ij}_\pm(\theta) = \frac{1}{2\pi}\int d\ell \ell J_{0/4}(\theta\ell)\left(C^{EE}_{ij}(\ell) \pm C^{BB}_{ij}(\ell)\right),
\end{equation}
where $J_{0/4}$ are the 0th / 4th-order Bessel functions of the first kind. In the curved sky limit, the cosmic shear TPCFs cannot be written as the Hankel transform of the power spectrum, and should be replaced by the spherical transform
\begin{align}
    \label{eq:xi_curved}
    \xi^{ij}_\pm(\theta) &= \sum_\ell\frac{2\ell+1}{4\pi}\left(C^{EE}_{ij}(\ell) \pm C^{BB}_{ij}(\ell)\right) \:{}_2d_{\ell, \pm2}(\theta) \\
    \label{eq:bkgd:hankel}
    &= H^\pm_{\theta, \ell}\left(C^{EE}_{ij}(\ell) \pm C^{BB}_{ij}(\ell)\right)
\end{align}
where ${}_{2}d_{\ell, \pm2}$ is the Wigner-d matrix and we use $H^\pm_{\theta, \ell}$ to denote the spherical transform operator. The summation is over all $\ell$ and consequently computing $\xi_\pm$ to an arbitrary accuracy can be computationally expensive. Here we have added the $B$ mode power spectrum for completeness, and equations~\eqref{eq:xi_curved} and~\eqref{eq:bkgd:hankel} apply to any spin-2 field. Hereafter, we will drop the subscript for tomographic bin pairs and write the multipole modes in the subscript instead.

\section{Method}
\label{sec:method}

In this section, we describe the measurement of the cosmic shear power spectrum (Sec.~\ref{sec:method:ps}) and two-point correlation functions (TPCFs; Sec.~\ref{sec:method:tpcf}) from the shear catalog. We discuss the methods to account for survey window and scale cuts, and how to consistently reconstruct the power spectrum from the TPCFs and vice versa in Sections~\ref{sec:method:tpcftops} and~\ref{sec:method:pstotpcf}. The motivation behind our method is to ensure \textit{matching} cosmic shear measurements between real-space and harmonic space analyses. Since each space has distinct advantages and challenges, properly accounting for survey masks, pixelization effects, and scale cuts is essential to avoid introducing systematic biases into cosmological inferences.


\subsection{Power Spectra}
\label{sec:method:PS}
In this subsection, we describe how the power spectra are measured in practice. In Sec.~\ref{sec:method:fullsky}, we describe the idealized case where we have a full-sky shear map. In Sec.~\ref{sec:method:partialsky}, we describe the biases that arise in power spectrum estimation due to partial sky and non-uniform coverage and introduce the formalism to correct it.

\subsubsection{Full Sky Shear Map}
\label{sec:method:fullsky}
For a full-sky uniform shear map, where the number density does not vary across the sky, the spin-2 continuous shear field can be decomposed into two scalar fields using spherical harmonics as
\begin{equation}
    \label{eq:method:shear_field}
    (\gamma_1 + i \gamma_2)(\bm{n}) = -\sum_{\ell m} \left[E_{\ell m} + i B_{\ell m}\right] {}_{\pm2} Y_{\ell m} (\bm{n}),
\end{equation}
where $\bm{n}$ is the sky position, $E_{\ell m}$ and $B_{\ell m}$ are the curl-free and divergence-free components of the shear field mentioned in Sec.~\ref{sec:bkgd:cosmic_shear_analysis}, and ${}_{s} Y_{\ell m}$ are the spin-weighted spherical harmonics. For two shear fields, the cosmic shear power spectrum is measured by taking the average of the product of two spherical harmonic coefficients $\psi_{\ell m}$ and $\phi_{\ell m}$ where $\psi, \phi \in \{E, B\}$ as
\begin{equation}
    \label{eq:method:cosmic_shear_ps}
    \langle \psi_{\ell m} \phi^*_{\ell' m'} \rangle = \delta_{\ell \ell'} \delta_{m m'} \hat{C}^{\psi \phi}_\ell.
\end{equation}
Throughout the paper, we will use $\langle \cdot \rangle$ to denote the ensemble average, $\Bar{x}$ to represent the sample mean of quantity $x$, and $\hat{x}$ to represent the measurement from data. In this work, we only consider the auto power spectra $C^{EE}_\ell$ and $C^{BB}_\ell$. For a full sky uniform coverage map, the optimal estimator of the cosmic shear power spectrum is then:
\begin{align}
    \label{eq:method:ideal_ee}
    \hat{C}^{EE}_\ell &= \frac{1}{2\ell+1} \sum_m E_{\ell m}E^*_{\ell m} \\
    \label{eq:method:ideal_bb}
    \hat{C}^{BB}_\ell &= \frac{1}{2\ell+1} \sum_m B_{\ell m}B^*_{\ell m},
\end{align}
where the power spectrum is averaged over the $m$ values for each multipole mode $\ell$. For notational convenience, we assume the two shear fields to be in the same redshift bin and omit explicit expressions for the tomographic redshift bin pairs for the rest of the paper. In the idealized case of a full sky, uniform survey, the power spectrum estimators are \textit{matching} between real and harmonic space because there is no survey mask or scale-dependent selection effects. However, real observations introduce non-trivial window functions and observational systematics that must be accounted for to ensure unbiased results.

\subsubsection{Masked Shear Map}
\label{sec:method:partialsky}
Equations~\eqref{eq:method:ideal_ee} and~\eqref{eq:method:ideal_bb} are for an idealized case -- they do not account for the survey geometry or for the scale cuts used in real space to remove scales that may be contaminated by observational systematics. In practice, we have non-uniform coverage of the sky, due to the limited exposure time, finite survey area, bad pixels, and regions masked due to bright stars. Once we obtain a galaxy shear catalog, we pixelize the shear value onto a finite grid. When using a pixelized map, the measurements can be affected by the pixel area and the number of pixels as the pixelized signal is the average within each pixel of the underlying true signal. The measured shear value at pixel $p$ is defined as the sum of the shear values of each galaxy in pixel $p$ divided by the mean number of galaxies: 
\begin{equation}
    \label{eq:method:pixel_shear_sum}
    \hat{\gamma}(\bm{n}_p) = \frac{1}{\bar{N}_g}\sum_{i\in p}{\hat{\gamma}_i},
\end{equation}
where $\bar{N}_g$ is the mean number of galaxies per pixel given by the ratio of total galaxy number to the number of pixels within the survey mask. The theoretical model for the pixelized shear field is given by
\begin{equation}
    \label{eq:method:pixel_shear_int}
    \gamma(\bm{n}_p)= \int \Pi(\bm{n} - \bm{n}_p) W^\gamma(\bm{n}) \gamma(\bm{n}) \mathrm{d}\bm{n},
\end{equation}
where $\bm{n}_p$ is the sky position of the $p$-th pixel, $\Pi$ is the pixel window function which is equal to the inverse of the pixel area for galaxies within the pixel and zero outside, 
and the weight $W^\gamma(\bm{n})$ (hereinafter the survey window) is defined by
\begin{equation}
    \label{eq:method:survey_window}
    W^\gamma(\bm{n}) = \frac{\hat{N}_g(\bm{n})}{\bar{N}_g},
\end{equation}
where $\hat{N}_g(\bm{n})$ is the observed number of galaxies at position $\bm{n}$. Note that $\bar{N}_g$ is a normalization choice we made and as long as normalization choice is treated consistently between data and model (including covariance), it does not have any impact on the analysis (its impact cancels out in the likelihood function). This definition includes clustering and shot noise in the survey window. Using a constant $\bar{N}_g$ in the denominator of equation~\eqref{eq:method:pixel_shear_sum} instead of the actual number of galaxies per pixel provides a more stable estimator of the shear. Dividing by the observed number of galaxies in each pixel would increases the noise in low-density pixels, as these estimators are highly sensitive to stochastic fluctuations in regions with fewer galaxies. Instead, the constant $\bar{N}_g$ acts as an implicit weighting factor; it reduces the relative contribution of noisy, low-density pixels while maintaining the contribution from high-density, low-noise pixels. Since this weighting is consistently included in the model, the final analysis remains unbiased. 
A more optimal choice of the window includes inverse variance weighting (the FKP weighting), which depends on the signal and noise power spectrum \citep{1994ApJ...426...23F, 1997MNRAS.289..285H},  but for this work, we adopt the weighting scheme of equation~\eqref{eq:method:survey_window}. In summary, equation~\eqref{eq:method:pixel_shear_sum} is the measured value of shear at pixel $p$, and equation~\eqref{eq:method:pixel_shear_int} is the theory shear value at pixel $p$ that accounts for the survey window and the pixel window function.

In equation~\eqref{eq:method:pixel_shear_int}, the pixelized shear field relates to the true underlying shear field via two operations in real space: multiplication of the survey window $W^\gamma$ followed by convolution of the pixel window function $\Pi$. Multiplication by the survey window $W^\gamma$ in real space is equivalent to convolving the shear harmonic coefficients in harmonic space. Measuring the power spectrum using the weighted shear field leads to a biased estimation as the survey window couples different $\ell$ modes and introduces $E$ mode contamination in $B$ modes and vice versa. 
Similarly, the convolution of the pixel window $\Pi$ and the weighted shear in real space corresponds to the harmonic coefficients of shear being multiplied by the Fourier transform of $\Pi$ in the harmonic space. This results in multiplying the power spectrum measurements in harmonic space by the square of the Fourier transform of $\Pi$. Since the pixel window function is localized in real space, the Fourier transform of $\Pi$ is unity for large scales $\ell \lesssim \pi/\Delta \theta$ and decreases for small scales $\ell \gtrsim \pi/\Delta\theta$, where $\Delta\theta$ is the angular resolution \citep{2014PhRvD..89b3003J}. This suppresses measurements of the power spectrum on smaller scales; we correct for this effect in the measurement by deconvolving the pixel window function in harmonic space. The measured power spectrum of this weighted map results in the pseudo-$C_\ell$ estimator ($D_\ell$) and is related to the true power spectrum via a coupling matrix \citep{2002ApJ...567....2H} as
\label{sec:method:ps}

\begin{align}
    \label{eq:pcl_ee}
    \hat{D}^{EE}_\ell &= C^{(\Pi)}_\ell\left( M^{+}_{\ell \ell'} C^{EE}_{\ell'}+M^{-}_{\ell \ell'} C^{BB}_{\ell'}\right)\\
    \label{eq:pcl_bb}
    \hat{D}^{BB}_\ell &= C^{(\Pi)}_\ell\left( M^{+}_{\ell \ell'} C^{BB}_{\ell'}+M^{-}_{\ell \ell'} C^{EE}_{\ell'}\right),
\end{align}
where $C^{(\Pi)}_\ell$ is the angular power spectrum of the spin-0 quantity $\Pi$ 
and $M^{\pm}_{\ell \ell'}$ is the coupling matrix. The spin-2 field coupling matrix is defined as
\begin{align}
    \label{eq:coup}
    M^{\pm}_{\ell \ell'} &= \frac{2\ell'+1}{4\pi} \sum_{\ell''} C^{(W^\gamma)}_{\ell''}(2\ell''+1) \left(\frac{1\pm(-1)^{\ell + \ell'+\ell''}}{2}\right) \\
    &\nonumber\times \begin{pmatrix}
        \ell & \ell' & \ell'' \\
        2 & -2 & 0
    \end{pmatrix} \begin{pmatrix}
        \ell & \ell' & \ell'' \\
        2 & -2 & 0
    \end{pmatrix}
\end{align}
where $C^{(W^\gamma)}_{\ell''}$ is the angular power spectrum of the survey window and $ \begin{pmatrix}
        \ell & \ell' & \ell'' \\
        m & m' & m''
    \end{pmatrix}$ is the Wigner 3-$j$ symbol (or Clebsch-Gordan coefficients). 
The $M^-_{\ell\ell'} C^{EE/BB}_{\ell'}$ terms correspond to the leakage introduced by the survey window. Throughout the paper, we use the Einstein summation convention to imply summation over repeated indices.

Additionally, one must account for the noise bias in the estimated power spectrum. The shape noise generates this bias due to the intrinsic ellipticities of galaxies. For a windowed field, the noise is effectively multiplied by the window given by $\sqrt{W^\gamma(\bm{n})}$ (see appendix~B of \cite{2020MNRAS.491...51S} for derivation of the window and the noise effects in pseudo-$C_\ell$ estimators). We analytically estimate the constant shape noise power spectrum as:
\begin{equation}
    \label{eq:shape_noise}
    D^{(N)}_\ell = M^N_{\ell \ell'} N_{\ell'} = \sum_{\ell'} M^N_{\ell \ell'}\frac{\sigma_e^2}{\bar{n}_\mathrm{eff}},
\end{equation}
where $\sigma_e$ is the shape noise, $\bar{n}_\mathrm{eff}$ is the effective number of galaxies in inverse steradians, and $M^N_{\ell\ell'}$ is the coupling matrix from using the window $\sqrt{W^\gamma(\bm{n})}$. 
We use the superscript $N$ to denote the coupling matrix for the noise in the shear field as the signal and noise have different windows. 
We will drop $C^{(\Pi)}_\ell$ and $N_\ell$ to simplify our notation, though we do correct for the pixel window effect and the additive noise bias on the measurements. 

In summary, this estimator introduces an artificial mixing of $E$- and $B$-modes due to the survey window. If uncorrected, this leads to systematic errors in cosmological parameter estimation.

\subsubsection{Accounting for Scale Cuts}

The pseudo-$C_\ell$ measurements in equations~\eqref{eq:pcl_ee} and \eqref{eq:pcl_bb} are not matching with traditional TPCFs measurements, and the two two-point estimators are weakly correlated \citep[e.g.,][]{2020PASJ...72...16H, 2021MNRAS.503.3796D}. This is because TPCFs are measured over a limited $\theta$ range in real space, but this is not explicitly modeled in the pseudo-$C_\ell$ measurements.  Therefore, the two measurements effectively use different sets of information that are probed by different $\ell$ ranges. As shown in \cite{2021MNRAS.508.1632S}, to make a matching two-point estimator in harmonic space with the TPCFs in real space, we need to account for the finite range of $\theta$ used in real space. Applying scale cuts in real space corresponds, in practice, to multiplying the TPCFs by a $\theta$-dependent function denoted as $\nu(\theta)$. Mathematically, this operation is represented as $\hat{\xi}_\pm(\theta) = \xi_\pm(\theta)\nu(\theta)$, where the scale cut $\nu(\theta)$ is typically a step function defined over a finite range of scales. 

Just as the product of the survey window in the shear map transformed into a convolution of the shear harmonic coefficients in the harmonic space, taking the spherical transform of TPCFs weighted by the scale cut $\nu(\theta)$ results in a convolution of power spectra in harmonic space. Since the pseudo-$C_\ell$ measurement in equations~\eqref{eq:pcl_ee} and~\eqref{eq:pcl_bb} do not include this effect, we manually convolve the measured $\hat{D}_\ell$ with a new coupling matrix, $M^{\pm,\nu}_{\ell\ell'}$, treating $\nu(\theta)$ as the window in equation~\eqref{eq:coup}, and define $F_\ell$ as the convolved pseudo-$C_\ell$. We name this new power spectrum as the consistent pseudo-$C_\ell$, a power spectrum obtained by convolving the pseudo-$C_\ell$ with a coupling matrix induced by imposing scale cuts in real space as \citep{2021MNRAS.508.1632S} 
\begin{align}
    \label{eq:fell_ee}
    \nonumber\hat{F}^{EE}_\ell &= M^{+,\nu}_{\ell\ell'} \hat{D}^{EE}_{\ell'} + M^{-, \nu}_{\ell\ell'} \hat{D}^{BB}_{\ell'}\\
    \nonumber&= \left(M^{+,\nu}_{\ell\ell'}M^{+,w}_{\ell'\ell''} + M^{-,\nu}_{\ell\ell'}M^{-,w}_{\ell'\ell''}\right)C^{EE}_{\ell''} \\
    \nonumber&+ \left(M^{+,\nu}_{\ell\ell'}M^{-,w}_{\ell'\ell''} + M^{-,\nu}_{\ell\ell'}M^{+,w}_{\ell'\ell''}\right)C^{BB}_{\ell''} \\
    &\equiv \mathcal{M}^+_{\ell\ell'}C^{EE}_{\ell'} + \mathcal{M}^-_{\ell\ell'}C^{BB}_{\ell'}
\end{align}
and similarly
\begin{equation}
    \label{eq:fell_bb}
    \hat{F}^{BB}_\ell = \mathcal{M}^+_{\ell\ell'}C^{BB}_{\ell'} + \mathcal{M}^-_{\ell\ell'}C^{EE}_{\ell'},
\end{equation}
where
\begin{equation}
    \label{eq:method:reduced_M}
    \mathcal{M}^\pm_{\ell \ell'} \equiv \left(M^{+,\nu}_{\ell\ell'}M^{\pm,w}_{\ell'\ell''} + M^{-,\nu}_{\ell\ell'}M^{\mp,w}_{\ell'\ell''}\right).
\end{equation}
We distinguish the coupling matrices using superscripts; superscript $w$ refers to the coupling matrix for the survey window $W^\gamma$ as in equation~\eqref{eq:coup} and $\nu$ refers to the coupling matrix with the scale cut function $\nu(\theta)$ acting as the window. The coupling matrix $\mathcal{M}$ is a linear combination of the product of the two coupling matrices, representing the combined impact of the two effects on the observed power spectrum. 
In short, equations~\eqref{eq:fell_ee} and~\eqref{eq:fell_bb} take into account both the survey window and the scale cut used in the real space. See Appendix~\ref{sec:appendix:B} for the derivation of $F_\ell$ from two-point correlation functions.

\begin{figure}
    \centering
    \includegraphics[width=0.5\textwidth]{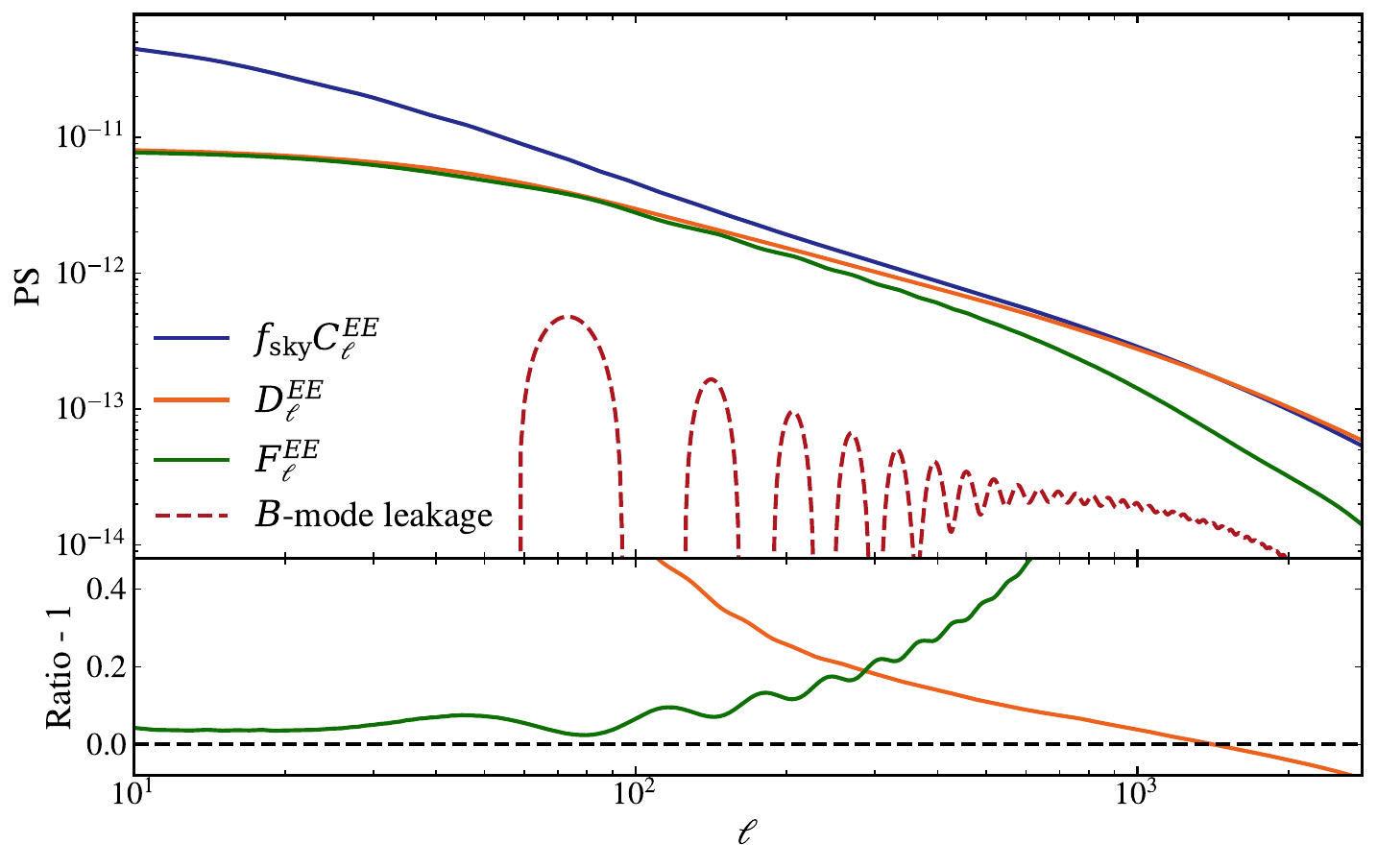}
    \caption{In the upper panel, we compare three different $E$-mode auto power spectra  for the lowest tomographic bin $(0.3, 0.6]$ of the fiducial cosmology defined in Sec.~\ref{sec:dai:mock}, using the HSC Y1 mask: $f_\mathrm{sky}C^{EE}_\ell$ (blue) with normalization convention of the pseudo-$C_\ell$ that includes a factor of $f_\mathrm{sky}$, $D^{EE}_\ell$ (orange), and $F^{EE}_\ell$ (green). The $B$-mode leakage of the form $M^{-,sc}M^{-,sw}C^{EE}_\ell$ in equation~\eqref{eq:method:reduced_M} is plotted as a dotted red line. We find that $B$-mode leakage contributes $\sim 5\%$ to the $E$-mode consistent pseudo-$C_\ell$. In the lower panel, we plot the ratios between the power spectrum and pseudo-$C_\ell$ (orange) and between pseudo-$C_\ell$  and consistent pseudo-$C_\ell$ (green). We find that information is lost on large scales when measuring the power spectrum on a weighted map, and power is recovered on small scales. Power is further lost on small scales when we model the impact of the scale cut used in real space in harmonic space.  
    }
    \label{fig:ps_comparison}
\end{figure}

In Fig.~\ref{fig:ps_comparison}, we compare three different $E$-mode model power spectra of the lowest redshift bin used in this work: the power spectrum $C^{EE}_\ell$ (blue), pseudo-$C_\ell$ $D^{EE}_\ell$ (orange), and consistent pseudo-$C_\ell$ $F^{EE}_\ell$ (green), and illustrate how each coupling matrix transforms one power spectrum to the other. We define the redshift bin used for this figure in Sec.~\ref{sec:dai:mock}. 
Going from the power spectrum to pseudo-$C_\ell$, we find that power is greatly suppressed on large scales (small $\ell$ values) by an order of magnitude and is recovered on small scales (large $\ell$ values) as can be seen from the lower panel ratio plot. This is because the weighted maps exclude (include) many realizations of small (large) $\ell-$modes. Comparing the orange and green curves, having a scale cut $\nu(\theta)$ in real space introduces further coupling between modes in harmonic space and has $\sim5\%$ effect on large scales and has $\gg 10\%$ effect for small scales for our adopted scale cut $10 < \theta < 360~[\mathrm{arcmin}]$ (see Sec.~\ref{sec:measurement:real_space} for details). 
Failure to account for this in harmonic space can bias power spectrum estimation on small scales. Scale cuts $\nu(\theta)$ are generally chosen to remove data points with significant evidence for observational systematic errors and/or model limitations. A smoother scale cut, as shown in equation~\eqref{eq:dai:wtheta} and illustrated in Fig.~\ref{fig:w(theta)}, is more optimal as it leads to convolution over fewer scales in $\ell$.  We also show the contribution of $B$-mode leakage into the $E$-mode of the consistent pseudo-$C_\ell$, which leads to errors of order a few percent. Reconstruction of power spectra without correction for such $B$-mode leakage can lead to biased results. Implementation details to produce Fig.~\ref{fig:ps_comparison} are discussed in Sec.s~\ref{sec:measurement:harmonic_space} and~\ref{sec:measurement:real_space}.

\subsubsection{Binning in $\ell$}

In general each coupling matrix $M_{\ell \ell'}$ is a rectangular matrix of order $O(\ell_\mathrm{max}^2)$ where $\ell_\mathrm{max} \sim 5000-7000$ or even higher. In order to reduce the dimensionality and effects of noise, we bin both the measurement $F_\ell$ and the product of coupling matrices $\mathcal{M}_{\ell \ell'}$. We use the notation convention in \cite{2021MNRAS.508.1632S} and define the binned coupling matrix as
\begin{align}
\label{eq:binned_fell_e}
    \nonumber F^{EE}_{\ell_b} &= \mathcal{B}^F_{\ell_b, \ell} F^{EE}_\ell \\
    &\nonumber= \mathcal{B}^F_{\ell_b, \ell} \mathcal{M}^{+}_{\ell \ell'} \left(B^{C^{EE}}_{\ell, \ell_b}\right)^{-1} \mathcal{B}^{C^{EE}}_{\ell'_b, \ell'} C^{EE}_{\ell'}\\
    &\nonumber+\mathcal{B}^F_{\ell_b, \ell} \mathcal{M}^{-}_{\ell \ell'} \left(\mathcal{B}^{C^{BB}}_{\ell, \ell_b}\right)^{-1} \mathcal{B}^{C^{BB}}_{\ell'_b, \ell'} C^{BB}_{\ell'} \\
    &= \mathcal{M}^{+EE}_{\ell_b \ell_b'} C^{EE}_{\ell_b'} + \mathcal{M}^{-BB}_{\ell_b \ell_b'} C^{BB}_{\ell_b'}
\end{align}
and similarly
\begin{align}
\label{eq:binned_fell_b}
    F^{BB}_{\ell_b} = \mathcal{M}^{+BB}_{\ell_b \ell_b'} C^{BB}_{\ell_b'} + \mathcal{M}^{-EE}_{\ell_b \ell_b'} C^{EE}_{\ell_b'},
\end{align}
where $\mathcal{B}$ is the binning operator defined in this work as
\begin{equation}
    \label{eq:method:binning_operator}
    \mathcal{B}_{\ell_b, \ell} = \begin{cases}
        \frac{2\ell+1}{(2\ell_b+1)\Delta \ell}, & \ell \in b \\
        0, & \mathrm{otherwise},
    \end{cases}
\end{equation}
where $\ell \in b$ is true when $\ell$ belongs to the bin centered on $\ell_b$ and $\Delta \ell$ is the bin size. The binned coupling matrices $\mathcal{M}_{\ell_b \ell'_b}$ with the superscript, e.g. $+EE$, refer to the power spectra used in defining the $\mathcal{B}$ operator. The same binning operator can be used for $F_\ell$ and $C_\ell$ as long as $\ell_\mathrm{max} \sim \ell'_\mathrm{max}$, so the coupling matrix is not too wide. We assume that the noise is estimated and subtracted from the measurements before binning. The binning operator in equation~\eqref{eq:method:binning_operator} is essentially weighting each $\ell$ mode by the effective number of modes. The choice of the binning operator is flexible and the $i$\textsc{Master} algorithm works with any binning operator as long as the same binning definition is used for all quantities. See \cite{2021MNRAS.508.1632S} for more optimal binning choices. 

Since binning data leads to loss of information, we generally cannot invert the binning operator. However, given a good model for the data, we can define the inverse of the binning operator as  \citep{2021MNRAS.508.1632S}
\begin{equation}
\label{eq:inverse_binning}
(\mathcal{B}^{C^{EE}}_{\ell, \ell_b})^{-1} = \begin{cases}
    \frac{C^{EE}_\ell}{C^{EE}_{\ell_b}} & \mathrm{if} ~\ell \in b\\
    0 & \mathrm{otherwise}
\end{cases},
\end{equation}
where $C^{EE}_{\ell_b} = \mathcal{B}^{C^{EE}}_{\ell, \ell_b} C^{EE}_\ell$ is obtained after binning the model $C^{EE}_\ell$. The same inverse binning operator can be used for different power spectra as long as the two power spectra have the same slope.

From $F^{EE}_{\ell_b}$ and $F^{BB}_{\ell_b}$ with equations~\eqref{eq:binned_fell_e} and~\eqref{eq:binned_fell_b}, we can reconstruct the underlying $E/B$-mode power spectrum as
\begin{align}
    \label{eq:rcell_fb_e}
    \hat{C}^{EE}_{\ell_b} &= \left(\mathcal{M}^{+EE}_{\ell_b \ell'_b}\right)^{-1}\hat{F}^{EE}_{\ell'_b} - \left(\mathcal{M}^{+EE}_{\ell_b \ell'_b}\right)^{-1}\mathcal{M}^{-BB}_{\ell'_b \ell''_b} C^{BB}_{\ell''_b} \\
    \label{eq:rcell_fb_b}
    \hat{C}^{BB}_{\ell_b} &= \left(\mathcal{M}^{+BB}_{\ell_b \ell'_b}\right)^{-1}\hat{F}^{BB}_{\ell'_b} - \left(\mathcal{M}^{+BB}_{\ell_b \ell'_b}\right)^{-1}\mathcal{M}^{-EE}_{\ell'_b \ell''_b} C^{EE}_{\ell''_b}.
\end{align}
In equations~\eqref{eq:rcell_fb_e} and~\eqref{eq:rcell_fb_b}, we also distinguished between the quantities measured from the data, denoted with $\:\hat{}$, and the ones estimated from theory. We are using the model $C^{EE/BB}_{\ell_b}$ to subtract out the leakage contribution. If the model is not well known and the noise is sub-dominant, we can use an iterative method where the initial estimates of $\hat{C}^{EE/BB}_{\ell_b}$ in equations~\eqref{eq:rcell_fb_e} and~\eqref{eq:rcell_fb_b} can be used to replace the model $C^{EE/BB}_{\ell_b}$ in the next iteration \citep[see discussion in Sec.~4.2.1 of][]{2021MNRAS.508.1632S}.
Hereafter, we will assume that the $B$-mode model power spectrum is zero, i.e. $C_{\ell}^{BB} = 0$, and consider only the $E$-mode theory to be nonzero. However, in practice, we still measure the pseudo-$B$-mode power spectrum induced by $E$-modes leaking into them. 

\subsection{Correlation Function}
\label{sec:method:tpcf}
In cosmic shear, the TPCFs measure the correlation of galaxy shapes given some angular separation $\theta$. The TPCFs can be measured from the shear catalog as
\begin{align}
    \label{eq:tpcf}
    \nonumber\hat{\xi}_\pm(\theta) &= \frac{\sum_{ij}\mathcal{W}(\bm{n}_i) \mathcal{W}(\bm{n}_j) \: \hat{\gamma}_{1}(\bm{n}_i) \hat{\gamma}_{1}(\bm{n}_j)}{\sum_{ij}\mathcal{W}(\bm{n}_i) \mathcal{W}(\bm{n}_j)} \\
    &\pm\frac{\sum_{ij}\mathcal{W}(\bm{n}_i) \mathcal{W}(\bm{n}_j) \: \hat{\gamma}_{2}(\bm{n}_i) \hat{\gamma}_{2}(\bm{n}_j)}{\sum_{ij}\mathcal{W}(\bm{n}_i) \mathcal{W}(\bm{n}_j)},
\end{align}
where the summation is over every galaxy pair $(i, j)$ with their angular separation $\theta = |\bm{n}_i - \bm{n}_j|$, 
$\mathcal{W}$ is the galaxy shape weight,  
and $\hat{\gamma}_1,\hat{\gamma}_2$ are the two measured shear components. 

Traditional TPCFs estimators return TPCFs that are average weighted by the number of pairs at each angular bin. 
However, the number of galaxies in each angular bin can be noisy with many bins, and the TPCFs estimator defined in equation~\eqref{eq:tpcf} does not fully remove the effects of the survey window. This is due to the effects of separately binning the numerator and denominator in equation~\eqref{eq:tpcf}. This way, the window effect is accounted for imperfectly, which leads to residual percent-level window effects when using traditional TPCFs estimators (see appendix~D of \citealt{2020MNRAS.491...51S} for demonstration of residual percent-level window effects on correlation functions). 
Rather than trying to use the above estimator and also identify and apply those additional window corrections, we take the alternate approach with a modified estimator from \cite{2021MNRAS.508.1632S} that is matching with the pseudo-$C_\ell$ 
defined in equations~\eqref{eq:pcl_ee} and~\eqref{eq:pcl_bb}, and is motivated by the weighting scheme in equation~\eqref{eq:method:survey_window}. 
Using an estimator that explicitly accounts for the survey window ensures \textit{matching} results with harmonic space analyses.

In this new TPCFs estimator, we divide the measurement from equation~\eqref{eq:tpcf} by the mean number of galaxies in each angular bin based on the survey window instead of the measured number of pairs, and use $\xi^{(\gamma)}_{W}(\theta)$ 
to denote the survey window correlation function
. Effectively, multiplying the TPCFs measurements with $\xi^{(\gamma)}_W(\theta)$ in real space corresponds to convolving the power spectrum with the coupling matrix in harmonic space, where the Hankel transform of $\xi^{(\gamma)}_W(\theta)$ is the angular power spectrum of the survey window in equation~\eqref{eq:coup}. We define the survey window correlation function as
\begin{equation}
    \label{eq:xi_w}
    \xi^{(\gamma)}_W(\theta) = \frac{\sum_{ij}\mathcal{W}(\bm{n}_i) \mathcal{W}(\bm{n}_j)}{f_\text{sky}N_\mathrm{source}^2\int_{\theta - \Delta\theta/2}^{\theta + \Delta\theta/2} 2\pi \sin\theta' d\theta'},
\end{equation}
where $f_\text{sky}$ is the fraction of the sky covered, $N_\mathrm{source}$ is the number of source galaxies in the tomographic bin,  
and $\Delta\theta$ is the width of the angular bins. The denominator in equation~\eqref{eq:xi_w} corresponds to the expected number of pairs in each bin if we have a full sky homogeneous survey window, rescaled by $f_\text{sky}$ to bring the normalization closer to the survey expectation (see Fig.~D1 of \citealt{2020MNRAS.491...51S} for an illustration of the survey window function). The numerator is based on weighted sums of pairs and the denominator is based on unweighted sums. However, as long as we use a self-consistent definition of the window in the measurements, model, and covariance, this inconsistency does not affect the results. 
Therefore, our new binned TPCFs are defined as \citep{2021MNRAS.508.1632S}
\begin{equation}
    \label{eq:method:xi_survey}
    \hat{\xi}_{\pm,W}(\theta) = \hat{\xi}_\pm(\theta)\xi^{(\gamma)}_W(\theta).
\end{equation}

In general, if possible, one should keep the data and model in the same space to avoid the complexities of transforming to the Fourier counterpart. Since the theory directly predicts the power spectrum in harmonic space, we reconstruct the cosmic shear power spectrum from the TPCFs measurements.


\subsection{Going from the TPCFs to the Power Spectrum}
\label{sec:method:tpcftops}
In this section, we describe how to consistently go from measured TPCFs in real space described in Sec.~\ref{sec:method:tpcf} to power spectrum in harmonic space. This transformation includes both the survey window and the scale cut used in both real and harmonic space.

As briefly mentioned in Sec.~\ref{sec:method:ps}, in practice, we measure the correlation function over a limited $\theta$ range as
\begin{equation}
    \label{eq:measured_xi}
    \hat{\xi}_{\pm, W, \mathrm{cut}}(\theta) = \hat{\xi}_{\pm,W}(\theta) \nu(\theta) = \hat{\xi}_\pm(\theta)\xi^{(\gamma)}_W(\theta)\nu(\theta),
\end{equation}
where $\xi^{(\gamma)}_W(\theta)$ is the correlation function of the window and $\nu(\theta)$ is the scale cut function, usually a top-hat function defined over $\theta \in [\theta_\mathrm{min}, \theta_\mathrm{max}]$. Multiplying the correlation function with the survey window correlation function and scale cut in real space results in convolution in harmonic space as \citep{2021MNRAS.508.1632S},
\begin{align}
    \label{eq:xitocl}
    \left(H^\pm_{\ell, \theta}\right)^{-1} ~\hat{\xi}_{\pm, W, \mathrm{cut}}(\theta) &= \mathcal{Q}^+_{\ell \ell'} C^{EE}_{\ell'} \pm \mathcal{Q}^-_{\ell \ell'} C^{BB}_{\ell'} = F^\pm_\ell
\end{align}
where $\mathcal{Q}_{\ell \ell'}$ is the coupling matrix with window power spectrum computed by taking the Fourier transform of the product $\xi^{(\gamma)}_W(\theta)\nu(\theta)$ in equation~\eqref{eq:coup}. Note that equation~\eqref{eq:xitocl} is the generalization of equation~\eqref{eq:bkgd:hankel}. In summary, to account for both the survey window and the scale cut multiplied in real space, one can convolve the power spectrum once with $Q^\pm_{\ell \ell'}$, or convolve the power spectrum twice with $\mathcal{M}^\pm_{\ell \ell'}$ in equation~\eqref{eq:method:reduced_M}.

Here we assume that $\hat{\xi}_+$ and $\hat{\xi}_-$ are measured over the same $\theta$ range. Note that when the $\theta$ range is small, its corresponding coupling matrix can be rather broad, requiring a larger $\ell$ range to make the power spectrum unbiased. To reduce the complexity of the coupling matrix, one can apodize the scale cut $\nu(\theta)$. \cite{2021MNRAS.508.1632S} shows a simple apodizing scheme where one can multiply the power spectrum of $\nu(\theta)$ with a cosine function that goes from one to zero over some $\ell$ range and transform back to obtain a new scale cut $\nu'(\theta)$ that brings some power from $\theta$ outside of its initial range and softens the edges of the top hat. In practical applications, the apodizing scheme will depend on the measurements being performed and the scales being used. 

We can further bin quantities similar to equations~\eqref{eq:binned_fell_e} and~\eqref{eq:binned_fell_b} and define a binned inverse spherical transform as
\begin{equation}
    \label{eq:method:binned_inverse_hankel}
    (H^\pm_{\theta_b, \ell_b})^{-1} =  \mathcal{B}_{\ell_b, \ell} (H^\pm_{\theta, \ell})^{-1} (\mathcal{B}^\pm_{\theta_b, \theta})^{-1},
\end{equation}
where the inverse binning operator $(\mathcal{B}^\pm_{\theta_b, \theta})^{-1}$ is defined similar to equation~\eqref{eq:inverse_binning} using the ratio $\xi_\pm(\theta)/\xi_\pm(\theta_b)$.
We can obtain the $E/B$-mode of $F_\ell$ as
\begin{align}
    \label{eq:method:epm_power}
    F^{EE}_{\ell_b} & = \frac{1}{2}(F_{\ell_b}^+ + F_{\ell_b}^-)\\
    \label{eq:method:bpm_power}
    F^{BB}_{\ell_b} & = \frac{1}{2}(F_{\ell_b}^+ - F_{\ell_b}^-).
\end{align}
Then, using equations~\eqref{eq:rcell_fb_e} and~\eqref{eq:rcell_fb_b}, we can reconstruct the underlying cosmic shear power spectrum. By accounting for the survey window and scale cuts, our method ensures that the reconstructed power spectrum is \textit{matching}, allowing for consistent cosmic shear analysis.


\subsection{Going from the Power Spectrum to the TPCFs}
\label{sec:method:pstotpcf}

Sometimes it is desired to transform the model defined in harmonic space into real space to compare with the data, as many cosmological models written in harmonic space have well-defined scale cuts. Similar to the issues we addressed in the case of TPCFs, multiplying the power spectrum with a scale cut in harmonic space should result in a convolution on the TPCFs in real space as
\begin{equation}
    \xi_{\pm, \ell-\mathrm{cut}} = H^\pm_{\theta, \ell} C_\ell b_\ell = b_{\theta \theta'}\xi_{\pm}(\theta'),
\end{equation}
where $b_\ell$ is the truncation function in harmonic space and $b_{\theta\theta'}$ is coupling matrix defined as
\begin{equation}
    b_{\theta\theta'} = \sum_\ell b_\ell \frac{2\ell+1}{4\pi} \:{}_{2} Y_{\ell, 2}(\theta) \:{}_{2} Y_{\ell, 2}(\theta').
\end{equation}

Similar to how we manually convolved the pseudo-$C_\ell$ to account for the scale cut used in real space, we can simply convolve the measured TPCFs with $b_{\theta\theta'}$ to impose the scale cut used in harmonic space before running the inference chain. By properly handling scale-dependent effects, we ensure that real space and harmonic space measurements provide \textit{matching} constraints on cosmological parameters.

\section{Data and Implementation}
\label{sec:dai}
In this section, we briefly introduce the HSC survey (Sec.~\ref{sec:dai:hsc}) and describe the mock catalogs used to test our new analysis methods (Sec. \ref{sec:dai:mock}). We discuss the details of the cosmic shear power spectrum measurements; we present the implementation details of pseudo power spectra in Sec.~\ref{sec:measurement:harmonic_space} and TPCFs in Sec.~\ref{sec:measurement:real_space}. We additionally describe the covariance matrix estimation in Sec.~\ref{sec:modeling_cov}.

\subsection{Hyper Suprime-Cam Survey}
\label{sec:dai:hsc}
The Hyper Suprime-Cam Subaru Strategic Program is a comprehensive imaging survey conducted using the 8.2~m Subaru Telescope and the Hyper Suprime-Cam wide-field camera (see \cite{2018PASJ...70S...4A} for survey design). The HSC survey consists of three layers: Wide, Deep, and UltraDeep. Each layer of the survey is observed in five broad-band filters ($grizy$) with a point source $5\sigma$ depth of $r \sim 26$ mag. The combination of depth and excellent seeing (typical $i$-band seeing of $\sim0.59$ arcseconds) allows HSC to measure cosmic shear signals up to higher redshifts with lower shape noise than other Stage III imaging surveys. 

The HSC Year 1 shear catalog (HSC-Y1) is based on data from the HSC S16A internal data release, consisting of data taken between March 2014 to April 2016 with about 90 nights. A number of cuts were made to construct a robust shape catalog for weak lensing analysis, including a conservative magnitude cut of $i < 24.5$, removing bad pixels, and masking regions of the sky around bright stars. The HSC-Y1 catalog covers a total area of $136.9\: \text{deg}^2$ that consists of six disjoint fields: XMM, GAMA09H, GAMA15H, HECTOMAP, VVDS, and WIDE12H (see Figure~1 in \citealt{2018PASJ...70S..25M} for more details). The shapes of galaxy are estimated on the $i$-band coadded images using the re-Gaussianization shape measurement method \citep{2003MNRAS.343..459H}. The method defines the two components of the galaxy ellipticity as
\begin{equation}
    (e_1, e_2) = \frac{1 - (b/a)^2}{1 + (b/a)^2} (\cos 2\phi, \sin 2\phi),
\end{equation}
where $b/a$ is the observed minor-to-major axis ratio and $\phi$ is the position angle of the major axis with respect to the equatorial coordinate system. The two components of the shear are then estimated using the measured ellipticity as
\begin{equation}
    \hat{\gamma}_\alpha = \frac{1}{2\mathcal{R}}\langle e_\alpha \rangle,
\end{equation}
where $\alpha=1, 2$ are the indices for the two spinor components and the shear responsivity $\mathcal{R}$ is the response of an ensemble of galaxies ellipticity to a small shear distortion \citep{1995ApJ...449..460K, 2002AJ....123..583B}. 

\subsection{HSC-Y1 Mock Shape Catalogs}
\label{sec:dai:mock}
In this work, we use the HSC-Y1 mock shape catalogs and verify that our new analysis method produces unbiased measurement of the underlying true tomographic cosmic shear power spectrum. These mock catalogs are generated following the
method described in \cite{Shirasaki_2019} using the full-sky ray-tracing
$N$-body simulations \citep{raytracingTakahashi2017}. In short, real HSC
galaxies are populated to the simulated lensing field according to their
measured positions on the transverse plane, and their redshift is randomly
drawn from the estimated \textsc{MLZ} photometric redshift posterior, where
\textsc{MLZ} is a machine-learning code based on a self-organizing map for
photometric redshift (photo-$z$) inference. Therefore, the mock catalogs have the same
survey geometry as the real HSC Y1 data. The six disjoint patches with holes in area coverage due to masking bright stars make it challenging to model the survey window $W^\gamma$ in equation~\eqref{eq:method:survey_window}, making the HSC-Y1 survey geometry a realistically challenging testbed for validating our new analysis method. 
As described in \cite{Shirasaki_2019}, the shear
catalog galaxies are divided into four tomographic redshift bins, in the
intervals $(0.3, 0.6], (0.6, 0.9], (0.9, 1.2],$ and $(1.2, 1.5]$, using the
``best" photo-z estimation by the \textsc{MLZ} algorithm.

The mock catalogs have galaxy positions in RA and Dec, true shear $\gamma_1$
and $\gamma_2$. We use the true shear value for our $F^{EE/BB}_\ell$ and $\xi_\pm$ measurements and use them to estimate the
covariance matrix as described in Sec.~\ref{sec:modeling_cov}. Since we use the true shear values for our measurements, the noise power spectrum $N_\ell$ in equation~\eqref{eq:shape_noise} is set to zero.


The mock catalogs adopt
a standard flat $\Lambda$CDM cosmological model with CDM density parameter
$\Omega_{c}=0.233$\,, the baryon density $\Omega_b=0.046$\,, the matter density
$\Omega_m=\Omega_c+\Omega_b=0.279$\,, the cosmological constant
$\Omega_\Lambda=0.721$\,, the Hubble parameter $h=0.7$\,, the amplitude of
density fluctuations $\sigma_8=0.82$\,, and the spectral index $n_s=0.97$. We refer to this as the fiducial model.

Before testing our method on the HSC-Y1 mock shape catalog, we first run our method on Gaussian mocks. We generate a set of Gaussian mocks using the \textsc{Healpy} \citep{2005ApJ...622..759G} package with the fiducial model power spectra and resolution parameter $N_\mathrm{side} = 8192$ corresponding to a pixel size of $\sim 0.43'$. Each Gaussian mock shares the same footprint as the HSC-Y1 mock shape catalog, allowing us to test our method for providing an unbiased reconstruction of the fiducial model.


\label{sec:measurement}

\subsection{Power Spectra From Pseudo Power Spectra}
\label{sec:measurement:harmonic_space}


We use the pseudo-$C_\ell$ method described in Sec.~\ref{sec:method:PS} to reconstruct the tomographic cosmic shear power spectra from the HSC Y1 shape mock catalogs. We measure the power spectra and mode coupling matrices to larger $\ell$ than we wish to use for our analysis to correctly account for mode coupling between smaller scale modes and to account for the $E/B$-mode leakage. The shear maps of each tomographic bin of the mocks were constructed using \textsc{Healpy} \citep{2005ApJ...622..759G} following equation~\eqref{eq:method:pixel_shear_sum} with a resolution parameter $N_\text{side} = 4096$, corresponding to a pixel size of $\sim 0.86'$. We measure the tomographic full-sky pseudo power spectra up to $\ell_\text{max} = 8000$, and the same $\ell_\text{max}$ was used to generate the coupling matrices $M^{\nu}_{\ell\ell'}$ and $M^{w}_{\ell\ell'}$ in equation~\eqref{eq:coup}. Before reconstructing the power spectra from the pseudo power spectra, we correct for the pixel window function by dividing the power spectra in equations~\eqref{eq:pcl_ee} and~\eqref{eq:pcl_bb} by the angular power spectrum of the pixel window function; $C^{(\Pi)}_\ell \simeq \operatorname{sinc}(\ell\Delta\theta/2\pi)$ is the pixel window function of \textsc{Healpix} \citep{2014PhRvD..89b3003J}, where $\Delta\theta$ is the angular resolution. We then convolve the pseudo power spectra with a coupling matrix derived from the scale cut (equations~\eqref{eq:fell_ee} and~\eqref{eq:fell_bb}) to estimate the consistent pseudo-$C_\ell$ and bin each spectrum and coupling matrices as described by the binning operator in equation~\eqref{eq:method:binning_operator}. Equations~\eqref{eq:rcell_fb_e} and~\eqref{eq:rcell_fb_b} are then used to reconstruct the $EE$ and $BB$ spectra for each tomographic bin pair.

We reconstruct $EE$ and $BB$ spectra in 20 linearly spaced multipole bins between $\ell_\text{min}=20$ and $\ell_\text{max}=5980$, 
and restrict to a narrower range and adapt the HSC Y1 analysis multiple ranges $300 < \ell < 1800$ \citep{2019PASJ...71...43H} when doing cosmological inference. Thus, this gives us 7 multipole bins that we use for our cosmological inference, from the 20 that we originally measure. For each mock, we have 10 spectra, four auto- and six cross-power spectra, and our data vector then consists of $10 \times 7 = 70$ data points.

\subsection{Power Spectra From TPCFs}
\label{sec:measurement:real_space}
\begin{figure}
    \centering
    \includegraphics[width=0.48\textwidth]{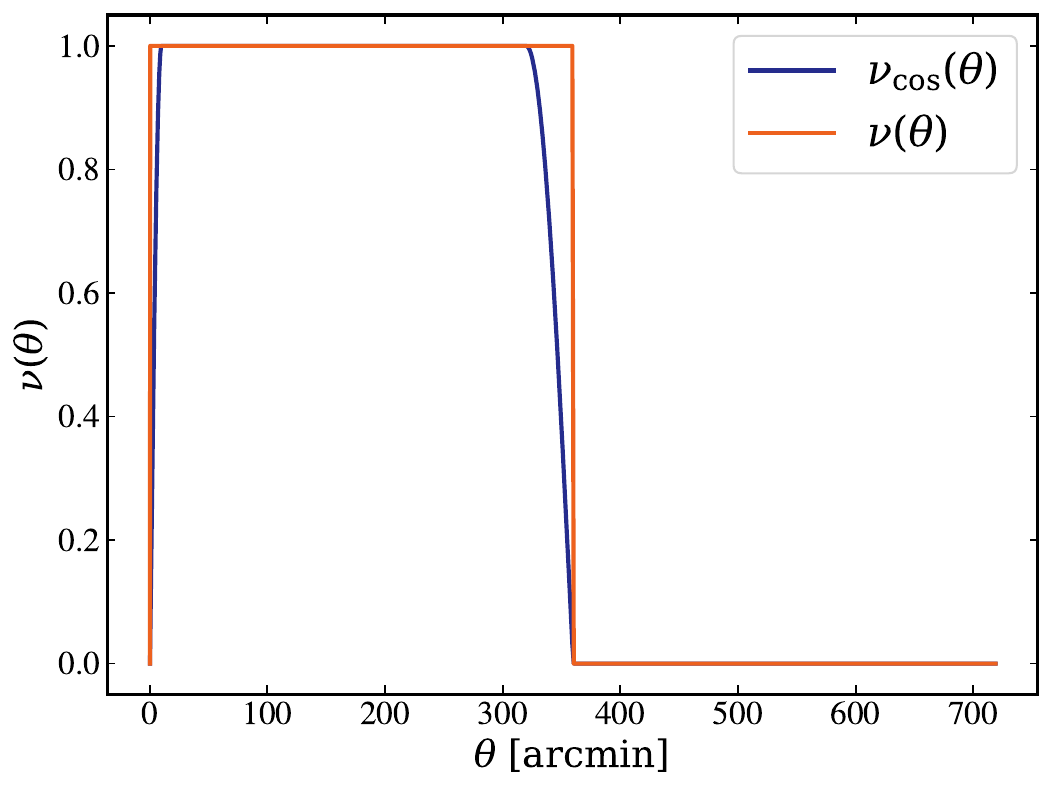}
    \caption{The scale cut imposed on the TPCFs measurements over angular scale. $\nu_{\cos}(\theta)$ (blue line) is 
 the tapered version of a flat top-hat (orange line) scale cut using equation~\eqref{eq:dai:wtheta} with $\theta_\text{cut,min}=10$ arcmin and $\theta_\text{cut,max}=320$ arcmin. The tapered version of the top-hat drops to zero more smoothly. For both scale cuts, the absolute minimum and maximum $\theta$ values are 0.25 and 360 arcmin. 
 }
    \label{fig:w(theta)}
\end{figure}
We use the public software \textsc{TreeCorr} \citep{TreeCorr}\footnote{\url{https://github.com/rmjarvis/TreeCorr}} to measure TPCFs. 
We use the TPCF method described in Sec.~\ref{sec:method:tpcf} to reconstruct the tomographic cosmic shear power spectra from the TPCF measurements. For each tomographic bin pair, we measure the TPCFs in 2400 logarithmically spaced angular bins between $0.25$ and $720$~arcmin. While $\nu(\theta)$ is typically chosen to be a top-hat function (default scale cut used for \textsc{Treecorr}), it can lead to convolution over rather large scales in $\ell$. We use the scale-cut function that drops more smoothly from one to zero (see Sec.~4.2.2 of \citealt{2021MNRAS.508.1632S})
\begin{align}
    \label{eq:dai:wtheta}
    \nonumber \nu_\text{cos}&(\theta,\theta_\text{cut,min},\theta_\text{cut,max}|\theta_\text{min},\theta_\text{max}) \\
    &= \begin{cases}
        0, &\theta \leq \theta_\text{min} ~\text{and}~ \theta \geq \theta_\text{max}\\
        \cos\left(\frac{\pi}{2}\frac{\theta-\theta_\text{cut,min}}{\theta_\text{cut,min}-\theta_\text{min}}\right), & \theta_\text{cut,min} < \theta < \theta_\text{cut,max}\\
        1, &\theta_\text{cut,min}\leq \theta \leq \theta_\text{cut,max} \\
        \cos\left(\frac{\pi}{2}\frac{\theta-\theta_\text{cut,max}}{\theta_\text{max}-\theta_\text{cut,max}}\right), & \theta_\text{cut,max} < \theta < \theta_\text{max}
    \end{cases},
\end{align}
where $\nu_\text{cos}(\theta)$ uses a cosine function to smoothly truncate the power spectra to zero between $\theta_\text{max} - \theta_\text{cut,max}$ and $\theta_\text{cut,min} - \theta_\text{min}$ with $\{\theta_\text{min}, \theta_\text{cut, min}, \theta_\text{cut, max}, \theta_\text{max}\} = \{0.25', 10', 320', 360'\}$ for the analysis. We first adopted the range of scales used in HSC Y1 and extended the range to reduce the complexity of the coupling matrix. A larger separation between $\theta_\text{cut,max}$ and $\theta_\text{cut,min}$ will lead to narrower convolution in $\ell$ space. The values of $\theta_{\text{cut,min}}$ and $\theta_{\text{cut,max}}$ were chosen as the Fourier transform of $\nu_{\cos}(\theta)$ drops to zero smoothly. We do not optimize $\nu_{\cos}(\theta)$ in this work. 
In Fig.~\ref{fig:w(theta)}, we compare the modified version of a top-hat scale cut with the aforementioned parameter values. The measured TPCFs are then multiplied with $\nu_{\cos}(\theta)$.

Once we have our TPCFs, we multiply by the survey window correlation function to match with the power spectra measurements, resulting in measurements as described in equation~\eqref{eq:measured_xi}. After binning the measurements and the inverse spherical transform operator, equations~\eqref{eq:method:epm_power} and~\eqref{eq:method:bpm_power} are used to get the pseudo-$D_\ell$, in which we reconstruct the power spectra using equations~\eqref{eq:rcell_fb_e} and~\eqref{eq:rcell_fb_b}. We reconstruct the $EE$ and $BB$ spectra in the same 20 linearly equal $\ell$ bins defined in \ref{sec:measurement:harmonic_space}, resulting in 70 data points for the data vector for each mock.

\subsection{Covariance Matrix}
\label{sec:modeling_cov}
We estimate the covariance matrix of our estimated cosmic shear power spectrum using mock catalogs. We do so by measuring the cosmic shear from $571$ realizations of the mocks and utilize these 571 to compute the covariance matrix.

The lensing covariance matrix can be decomposed into three terms: the Gaussian, connected non-Gaussian, and the super-sample covariance (\citealt{2018JCAP...10..053B}). The non-Gaussian terms are not modeled as the covariance matrix estimation is done on Gaussian mocks. The Gaussian covariance term of the cosmic shear power spectrum can be further decomposed into three terms: the auto-term of cosmic shear, the cross-term of cosmic shear and shape noise spectra, and the auto-term of the shape noise spectra as
\begin{equation}
\label{eq:gaussian_cov}
    \operatorname{Cov}^{(G)}(\hat{C}^{ij}_{\ell_1}, \hat{C}^{mn}_{\ell_2}) = \operatorname{Cov}^{(G)}_{SS} + \operatorname{Cov}^{(G)}_{SN} + \operatorname{Cov}^{(G)}_{NN},
\end{equation}
where $G$ denotes the Gaussian part of the covariance, $i,j,m,n$ are the tomographic bins, $S$ denotes the cosmic shear, and $N$ denotes the shape noise. Since we make the measurements on noiseless data, our estimation of the covariance from noiseless measurements only estimates the auto-term of the cosmic shear in equation~\eqref{eq:gaussian_cov}. Neglecting the shape noise contribution to the covariance can lead to adding more weights to small scales, which are generally more biased from the model. 

To get the full Gaussian covariance, we model the cross-term between cosmic shear and shape noise spectra and the auto-term of the shape noise spectra as
\begin{align}
    \label{eq:sncov}
    \operatorname{Cov}^{(G)}_{SN}(\hat{C}_{\ell_1}^{ij}, \hat{C}_{\ell_2}^{mn}) &= \frac{\delta_{\ell_1\ell_2}}{f_{\text{sky}}(2\ell_1+1)\Delta \ell_1}\left[\frac{\sigma_e^2}{\Bar{n}_\text{eff}^i}\left(C_{\ell_1}^{jn}\delta_{im} + C_{\ell_1}^{jm}\delta_{in}\right)\right.\\
    &\nonumber\left.+\frac{\sigma_e^2}{\Bar{n}_\text{eff}^j}\left(C_{\ell_1}^{im}\delta_{jn} + C_{\ell_1}^{in}\delta_{jm}\right)\right]\\
    \operatorname{Cov}^{(G)}_{NN}(\hat{C}_{\ell_1}^{ij}, \hat{C}_{\ell_2}^{mn}) &=\frac{\delta_{\ell_1\ell_2}}{f_{\text{sky}}(2\ell_1+1)\Delta \ell_1}\frac{\sigma_e^4}{\Bar{n}_\text{eff}^i\Bar{n}_\text{eff}^j}\left(\delta_{im}\delta_{jn} + \delta_{in}\delta_{jm}\right),
\end{align}
where $\sigma_e = 0.25$ is the RMS ellipticity of the source galaxies and we use values of  $\Bar{n}_\text{eff}^i$ from Table 1 column $n^{(C13)}_{g,\text{eff}}$ in \cite{2019PASJ...71...43H}. The Kronecker deltas $\delta$ ensure that the Gaussian covariance is non-vanishing when $\ell_1$ and $\ell_2$ belong in the same $\ell$ bin and the shape noise spectra contribute only to matching tomographic bins. Note that our analysis uses the true lensing shear values in the mock catalogs, so there is no shape noise in the data vectors. However, we use the above covariance (including the cross-term between cosmic shear and shape noise and the auto-term of the shape noise) so that the relative weight given to the bins in $\ell$ in our analysis is similar to what would be used in a realistic case with real data. By applying this covariance to the data vectors without shape noise, we obtain a precise estimate of any biases due to our method of power spectrum reconstruction with a realistic weighting across $\ell$ bins.


\begin{figure*}
    \centering
    \includegraphics[width=\textwidth]{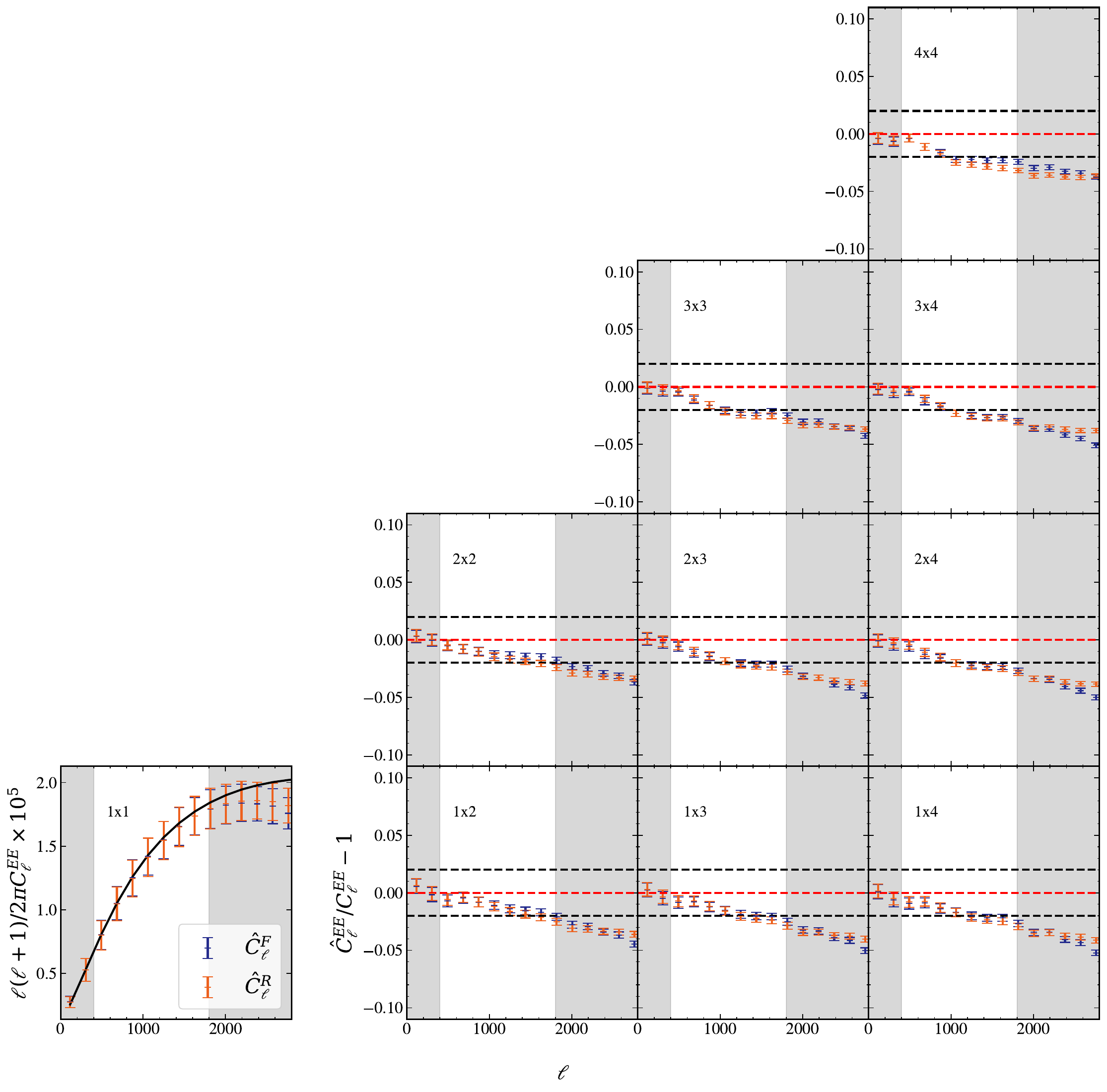}
    \caption{Comparison of the reconstructed tomographic cosmic shear power spectra of $EE$ mode from pseudo power spectra and TPCFs measurements with error bars in the lower left panel. The colored points show the average reconstructed power spectrum from 571 mock realizations from pseudo power spectra (blue dots) and two-point correlation functions (orange dots) for the lowest redshift bin, as indicated by the $1\times 1$ label. The analytical forward-modeled prediction (solid black line) is also shown for comparison. The scales outside of $400 < \ell < 1800$ (shaded regions) are excluded from the inference. The other panels show the ratio of each reconstruction to the forward-modeled theory, with error bars. The error bars are $1\sigma$ uncertainties in the ratio estimated using mock catalogs. 
    Reconstructions of power spectra from $i$\textsc{Master} give matching results to well within $1\%$ and they give unbiased results to within $2\%$. 
 Black dotted lines represent the $2\%$ bias region. 
 The scales outside of the region $400 < \ell < 1800$ (shaded regions) are excluded from the cosmological analysis. 
    }
    \label{fig:reconstruction}
\end{figure*}

\section{result}
\label{sec:result}


In this section, we discuss the result of our analysis of HSC Y1 mock shape catalogs described in Sec.~\ref{sec:dai:mock}. In Sec.~\ref{sec:result:reconstruction}, we show the result of power spectrum reconstruction from consistent pseudo-$C_\ell$ and TPCFs measurements. In Sec.~\ref{sec:result:likelihood}, we describe the outcome of fitting a cosmological model to the reconstructed power spectrum and performing cosmological parameter inference. 
In Sec.~\ref{sec:result:scale}, we discuss the scalability of the $i$\textsc{Master} algorithm. 

\subsection{Cosmic Shear Power Spectrum Reconstruction}
\label{sec:result:reconstruction}
Here, we present the reconstruction of the cosmic shear power spectrum from two-point function measurements in both real and harmonic space over multipole ranges specified in Sec.s~\ref{sec:measurement:harmonic_space} and~\ref{sec:measurement:real_space}.

Before running our method on the HSC mocks, we first test the accuracy of the reconstruction method on Gaussian simulations with an HSC footprint as described in Sec.~\ref{sec:dai:mock}. 
Our analysis method on the simple Gaussian simulations gives unbiased reconstructions of the power spectrum (one from the consistent pseudo-$C_\ell$ and the other from TPCFs), and the two reconstructions agree within $1\%$ for $300 < \ell < 1800$. 

Running the method ignoring the second term in equations~\eqref{eq:binned_fell_e} and~\eqref{eq:binned_fell_b}, or ignoring the leakage of the $B$ mode when constructing $F_\ell$, gives a biased reconstruction of the power spectrum and loss of the signal by $15-20\%$ for our survey window and scale cuts. 
Testing our method on Gaussian mocks, we conclude that our method gives unbiased reconstruction of the power spectrum on a survey geometry as complex as HSC Y1, and that $B$-mode leakage correction is needed given the scale cuts imposed in HSC.


In Fig.~\ref{fig:reconstruction}, we show the mean of the tomographic cosmic shear power spectrum reconstructed from both consistent pseudo-$C_\ell$ (blue dots) and TPCFs measurements (orange dots) performed on HSC mocks with error bars. We show the auto-power spectra for the lowest redshift bin and show the ratio of the reconstruction to the forward-modeled theory for the rest.
The two reconstructions agree within $\pm1\%$ with each other and within $\pm2\%$ with the forward-modeled theory in the $\ell$ ranges used in the analysis. However, the two reconstructions start to deviate from the fiducial model at higher $\ell$ values; fixing $\ell_\text{max}$ when measuring pseudo-$C_\ell$ power spectra and the coupling matrices can miss some of the power leakage across scales, biasing the reconstructed power spectra. This effect is largest for higher $\ell$ as those scales miss the most leakage. These biases can be reduced by increasing $\ell_\text{max}$, with increased computational cost. There may also be some bias from the imperfect power spectra model used in the reconstruction, and these biases can be reduced by decreasing the size of the bin \citep[see discussion in][]{2021MNRAS.508.1632S}.

\begin{figure*}
    \centering
    \includegraphics[width=0.8\textwidth]{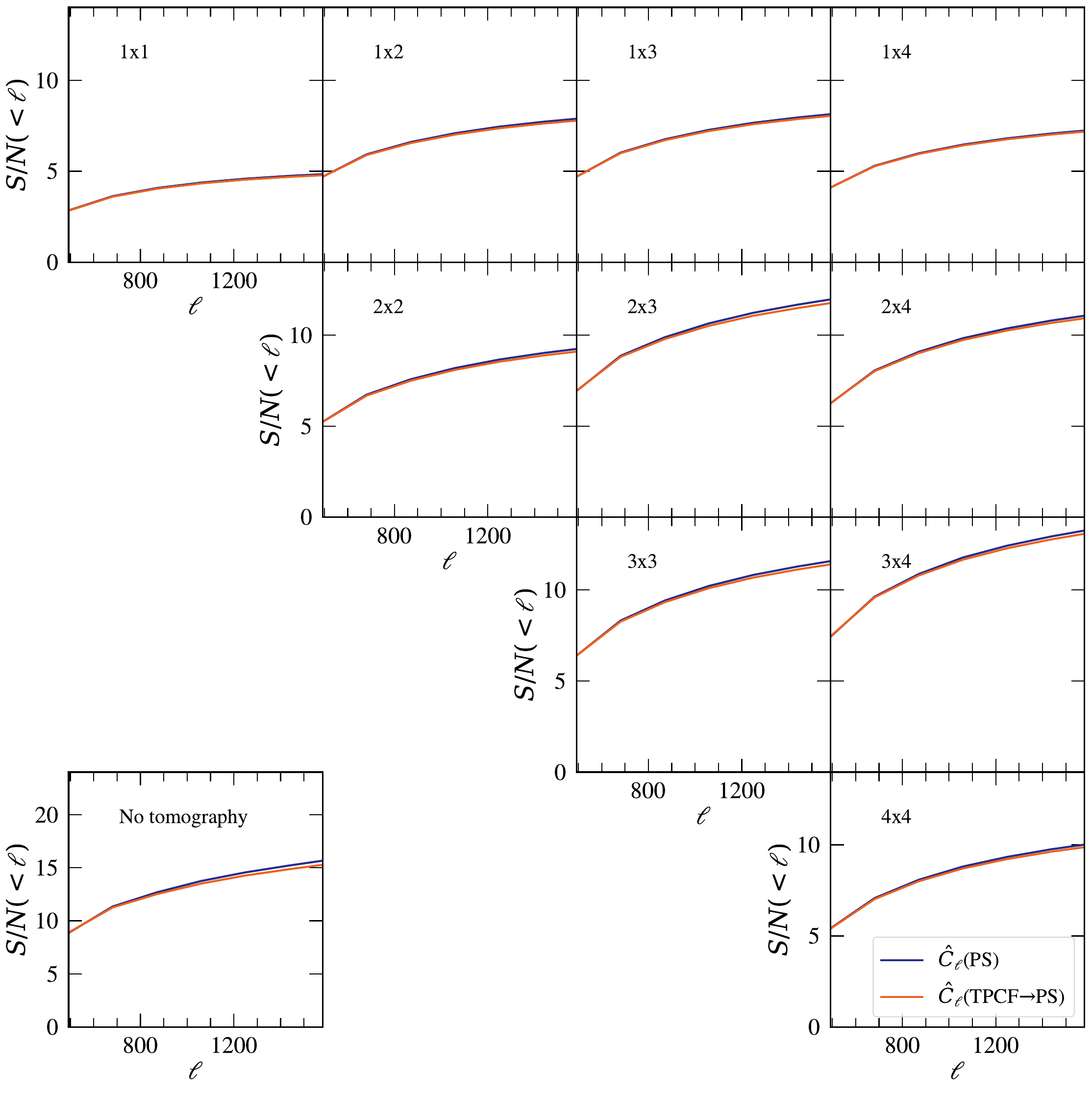}
    \caption{The integrated S/N of the reconstructed auto and cross power spectra from consistent pseudo-$C_\ell$ (blue) and two-point correlation function (orange) measurements as a function of $\ell$. The left-bottom panel shows the non-tomographic S/N of the reconstructed power spectrum. The two integrated S/N of the reconstructions are in agreement within $\pm1\%$. The S/N in these redshift bins are consistent with those of the HSC Y1 auto-power spectrum to within $20\%$. }
    \label{fig:snr}
\end{figure*}

In Fig.~\ref{fig:snr}, we show the comparison of the integrated signal-to-noise ratio (S/N) for tomographic cosmic shear up to some $\ell$ value. 
The S/N is defined as
\begin{equation}
    \label{eq:result:snr}
    S/N(q) = \sqrt{\hat{C}_\ell(q) \Sigma^{-1}(q) \hat{C}^T_\ell(q)},
\end{equation}
where $q \in \{\text{PS}, \text{TPCF}\to\text{PS}\}$ represents the direction of the analysis and $\Sigma$ is the covariance matrix estimated from the mocks. The two integrated S/N of the reconstructions are in agreement within $\pm1\%$. The S/N of the reconstructed power spectrum from the consistent pseudo-$C_\ell$ in individual redshift bins are 4.86, 9.32, 11.68, and 10.09 from the lowest to highest redshift bins, respectively. The S/N in these redshift bins are consistent with those of the HSC Y1 power spectrum and correlation function analyses to within $20\%$. The S/N from mocks is expected to be less than that of real data, as real data contains noise that artificially biases the measured signal to noise ratio to higher values. To quantify the covariance of two reconstructions, we compute the correlation coefficient as
\begin{equation}
    \label{eq:result:corrcoef}
    r(x) = \frac{\operatorname{Cov}(x_R, x_F)}{\operatorname{Cov}(x_R, x_R )^{1/2}\operatorname{Cov}(x_F, x_F)^{1/2}},
\end{equation}
with $x = \hat{C}_\ell$. The subscripts $R$ and $F$ stand for the reconstruction from TPCFs (real space) and consistent pseudo-$C_\ell$ (harmonic space), respectively. We find that the correlation coefficient range is $r(\hat{C}_\ell)\in[0.91, 1.0]$, thus the two reconstructions have the same information and are matching. For a detailed review, refer to Appendix~\ref{sec:appendix:A}.  

\subsection{Likelihood Inference}
\label{sec:result:likelihood}
Once we obtain measurements of the cosmic shear power spectrum using the method described in Sec.s~\ref{sec:measurement:harmonic_space} and~\ref{sec:measurement:real_space}, the next stage of the analysis is the likelihood analysis. In this step, the tomographic cosmic shear power spectrum data vector is compared with a theoretical data vector computed using a forward model based on cosmological parameters. 

To obtain a single parameter quantifying the fidelity of power spectrum reconstruction, we constrain the cosmological parameter $S_8$ by comparing the reconstructed power spectrum data vector measured in Sec.s~\ref{sec:measurement:harmonic_space} and~\ref{sec:measurement:real_space} to model-predicted power spectrum. A higher-dimensional parameter inference is possible with this new estimator and is left for future work; we deliberately simplify in this case so the interpretation in terms of biases in the cosmic shear power spectrum amplitude is unambiguous. When computing the theory cosmic shear power spectrum in equation~\eqref{eq:bkgd:ps}, we use the publicly available code \textsc{CAMB} \citep{2011ascl.soft02026L}\footnote{\url{https://github.com/cmbant/CAMB}} to compute the linear part of the matter power spectrum and use \textsc{pyhalofit}\footnote{\url{https://github.com/git-sunao/pyhalofit}} to compute the non-linear matter power spectrum. 
To properly account for the shell thickness effect and finite resolution effect in the full-sky ray-tracing simulations of the mock catalog, we modify the matter power spectrum as described in \cite{Shirasaki_2019}. We then use the Limber approximation following equation~\eqref{eq:bkgd:ps} to get the tomographic cosmic shear power spectra with the matter power spectrum. After obtaining the tomographic cosmic shear power spectra, 
we forward model them using the methods described in Secs.~\ref{sec:method:PS} and~\ref{sec:method:tpcf} to incorporate the effects of the survey window and scale cut. 

For each realization of the mock, we define the $\chi^2$ of our model in terms of the Gaussian log-likelihood as:
\begin{align}
    \nonumber \chi^2(q|\Theta) &= -2\log \mathcal{L}(\hat{C}_{\ell_b}(q) | \Theta)\\
    &= \left(\hat{C}_{\ell_b}(q) - \Tilde{C}_{\ell_b}(q|\Theta)\right)^T \Sigma^{-1} \left(\hat{C}_{\ell_b}(q) - \Tilde{C}_{\ell_b}(q|\Theta)\right),
\end{align}
where $\chi^2$ comes from the concatenated tomographic cosmic shear power spectra data vector and $q\in\left \{\text{PS}, \text{TPCF}\to\text{PS}\right \}$. $\hat{C}_{\ell_b}$ is the reconstructed power spectrum from the mocks and $\Tilde{C}_{\ell_b}$ is obtained after forward modeling the model cosmic shear power spectrum. We ignore the $B$-mode power spectrum during the inference as our reconstructed and model $B$-mode are both zero, i.e. $C^{BB}_\ell = \hat{C}^{BB}_\ell = 0$. 

\begin{figure*}
    \centering
    \includegraphics[width=\textwidth]{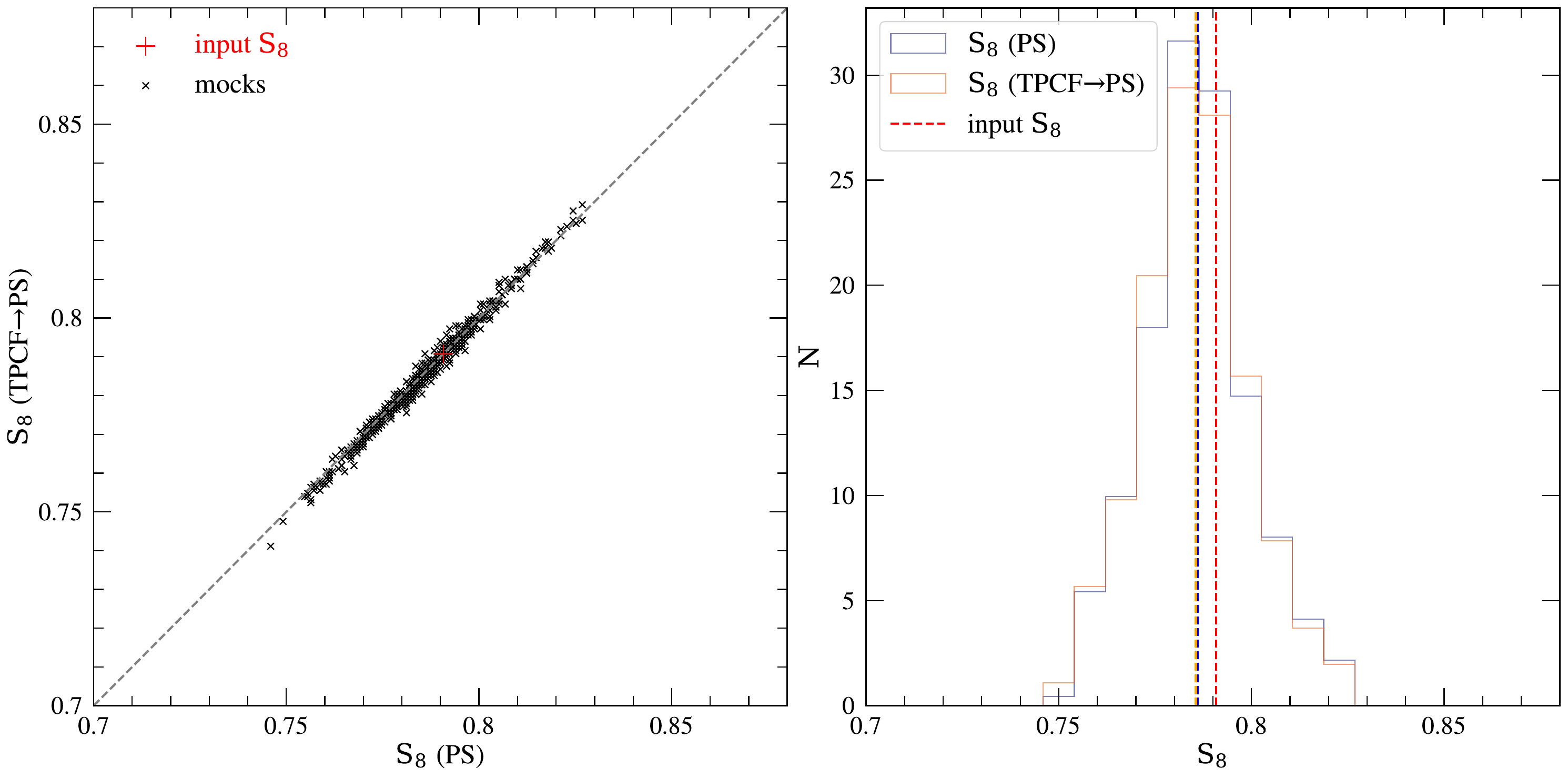}
    \caption{Left panel: scatter plot showing the inferred $S_8$ values from consistent pseudo-$C_\ell$ analysis in the horizontal axis and from two-point correlation functions analysis in the vertical axis. The points are from minimizing $\chi^2$ for 571 mock catalogs. The red $+$ denotes the value of $S_8$ adopted in generating the mock catalogs. The inferred $S_8$ values between the two independent analyses are very consistent, with the correlation coefficient $r=0.994$. While the comparison is conceptually similar to Figure~17 of \citep{2020PASJ...72...16H}, which highlights the inconsistency between different summary statistics, we emphasize that their analysis was performed on noisy mock data, whereas our analysis uses noiseless simulations. As such, the results are not directly comparable. Right panel: we show the distribution of $S_8$ from the mocks for each axes. The dotted vertical lines represent the mean of each distribution, with the red dotted vertical line denoting the fiducial $S_8$ value. The two distributions have very similar means, with $\Delta \langle S_8\rangle = 0.0004$.}
    \label{fig:s8inference}
\end{figure*}

Motivated by Figure~17 of \cite{2020PASJ...72...16H}, we use a simple grid-based parameter estimation method to infer $S_8$, starting from both real and harmonic space measurements and reconstructing the power spectrum. We precompute the model power spectrum from a grid space of parameters of $\Theta = \{S_8\}$ and $S_8 \in [0.4, 1.2]$ with a bin size of $\Delta S_8 = 0.008$. Then, for each mock catalog, we find the value of $S_8$ on the grid that maximizes the log-likelihood or minimizes $\chi^2$ as:
\begin{equation}
    S^i_8(q) = \underset{S_8\in[0.4, 1.2]}{\arg\min} \chi^2_i(q|S_8),
\end{equation}
where $S^i_8$ is the inferred $S_8$ value using the $i$th mock catalog. We compare the inferred $S_8$ values using the reconstructed cosmic shear power spectrum from consistent pseudo-$C_\ell$ and TPCFs measurements on the same 571 mock catalogs in the scatter plot in Fig.~\ref{fig:s8inference}. 

We use equation~\eqref{eq:result:corrcoef} to quantify the covariance of the two inferred $S_8$ values, and we find that $r(S_8) = 0.994$, 
which confirms that the two $S_8$ values are strongly correlated. \cite{2020PASJ...72...16H} did a similar test in which they inferred two parameters ($S_8$ and $\Omega_m$) separately from TPCFs and power spectrum measurements. The difference between their analysis and our analysis is that we reconstructed the power spectrum from the TPCF measurements when running the inference. They find only weak correlations between the inferred cosmological parameters from the two analyses: $r(S_8) = 0.51$ and $r(\Omega_m) = 0.17$. The reason for such disparity between the two analyses is due to the different multipole ranges probed in the analyses; they showed that the scale cut $\nu(\theta)$ used in the HSC Y1 analysis has a large contribution from scales with $\ell < 300$, which were excluded in the pseudo-$C_\ell$ analysis, leading to the two analyses using different and complementary information. A direct comparison of the two $r(S_8)$ values is challenging as \cite{2020PASJ...72...16H} performed a multidimensional parameter inference while marginalizing over nuisance parameters. We show that we get a consistent parameter inference once we account for this scale cut used in real space in the consistent pseudo-$C_\ell$ measurement. 

The right panel in Fig.~\ref{fig:s8inference} shows the distribution of inferred $S_8$ values from reconstruction using consistent pseudo-$C_\ell$ (blue) and TPCFs (orange) measurements, with the red dashed line showing the fiducial value of $S_8$. The mean of each distribution is $0.7860$ and $0.7856$ with standard deviations of $0.0137$ and $0.0143$, respectively. The difference between the mean and median values of the inferred values of $S_8$, $\delta S_8$, are $0.0004$ and $0$, respectively. This difference is less than the observed difference of $-0.042$ for HSC Y1 by two orders of magnitude. The means of both inferred $S_8$ values are consistent with the fiducial value to well within $0.7\%$. Our results suggest that the method as implemented here can successfully provide a self-consistent representation of the weak lensing power spectrum starting from pseudo-$C_\ell$ in harmonic space or from the two-point correlation functions in real space, and residual subpercent biases (consistent across both methods) may be due to factors outside of this method. One possible source of residual biases is the imperfect correction for simulation resolution effects in the mock catalogs.  


\subsection{Scalability of $i$\textsc{Master}}
\label{sec:result:scale}
One of the advantages of using $i$\textsc{Master} method is that it speeds up the computation during sampling by reducing the complexity of the coupling matrix to $O(N^2_\text{bin})$ instead of $O(\ell^2_\text{max})$ by binning quantities, where $N_\text{bin} \sim O(10)$. Assuming that the survey window is fixed, we only have to compute the coupling matrices $M^{\pm, w}_{\ell \ell'}$ and $M^{\pm, \nu}_{\ell \ell'}$ once
, and they are used to get the consistent pseudo-$C_\ell$. Since survey window estimation may have uncertainties, coupling matrices may need to be recomputed for multiple windows \citep[see ][]{2021MNRAS.508.1632S,Karim2023}. Since we propose to convolve the pseudo-$C_\ell$ with a coupling matrix from scale-cuts in real space, the complexity becomes $O(N^3_\text{bin})$ from the matrix multiplication. 
\section{Conclusion}
\label{sec:conclusion}


In this paper, we presented the first practical application of the $i$\textsc{Master} algorithm \citep{2021MNRAS.508.1632S} to mock catalogs with a realistic cosmological density field and a realistically complex survey window (the HSC Y1 mock shape catalogs).  In doing so, we reviewed the formalism for the measurement of cosmic shear two-point functions and how to correctly account for survey window and scale cuts in harmonic and real space to obtain a \textit{matching} power spectrum. 

In Sec.~\ref{sec:method}, we discussed the impact of the survey window and scale cut on the various estimators. Applying the survey window and scale cuts in real space generally involves multiplying the underlying maps by some weights, which are equivalent to convolutions in harmonic space. Since measurements of the power spectrum with non-uniform coverage of the sky do not account for the scale cut used in real space, we manually convolved the pseudo-$C_\ell$ to obtain the consistent pseudo-$C_\ell$. We showed that this greatly affects the pseudo-$C_\ell$ measurements in all scales. We also showed that leakage of the form $M^{-,sc}M^{-,sw}C^{EE}_\ell$, equivalent to $B$-mode leakage of $F^{EE}_\ell$, contributes $5$ per cent to the $F^{EE}_\ell$ measurements. We also derived a new two-point correlation function (TPCF) estimator that is matching with the process of estimating the pseudo-$C_\ell$.  In particular, the new estimator normalizes the measurements by the expected number of galaxy pairs instead observed number of galaxy pairs, which is traditionally done in the literature. This new procedure changes the two-point correlation function measurements by $\sim$2 per cent and allows for self-consistent modeling in the two spaces. 

We also showed methods to reconstruct the $E/B$ mode power spectra from pseudo power spectra and two-point correlation function measurements. In this paper, we assumed that $\xi_\pm$ are measured over the same range of scales. If one wishes to use different scales for $\xi_+$ and $\xi_-$, one must generalize equations~\eqref{eq:method:epm_power} and \eqref{eq:method:bpm_power}. 

In Sec.~\ref{sec:dai}, we discussed the HSC Y1 shape mock catalogs used for the analysis and the implementation details of the cosmic shear power spectrum measurements. The reconstruction of the power spectra from TPCFs measurements can be challenging if the range of scales at which TPCFs are measured is limited, leading to a complex coupling matrix. We introduced a new scale-cut that drops to zero smoothly; this smoothed scale-cut would lead to a narrower convolution in $\ell$ space and reduce the complexity of the coupling matrix. However, we do not optimize the scale-cut in this work.

In Sec.~\ref{sec:result}, we showed the result of applying the $i$\textsc{Master} algorithm using the HSC Y1 shape mock catalog. Repeating the scale cuts used in both harmonic and real space in the actual HSC Y1 analysis, $i$\textsc{Master} algorithm was able to reconstruct the cosmic shear power spectrum from $F^{EE/BB}_\ell$ and $\xi_\pm$ that match within $\pm1$ per cent with each other. We also showed that the two integrated S/N of the reconstructions are matching within $\pm1$ per cent. We compared the inferred $S_8$ values using likelihood inference with the reconstructed cosmic shear power spectrum. We found that the two inferred $S_8$ values from 571 mocks to be strongly correlated with a correlation coefficient of 0.994. 

This algorithm has more capabilities than existing algorithms (e.g., the \textsc{Master} and \textsc{NaMaster} algorithms) as it enables direct comparisons with unbinned theoretical models calculated at effective bin centers. Additionally, our algorithm is efficient and consistent in that it allows for more precise extraction of information for a specific set of scales by effectively removing the mode mixing in the consistent pseudo-$C_\ell$ estimator, and we expect it will be useful for cosmic shear data analysis.

    


The $i$\textsc{Master} implementation provides options to speed up computations after the initial setup (computing coupling matrices). The computational complexity is reduced from $O(\ell_\text{max}^2)$ to $O(N_\text{bin}^2)$, and similar scaling for computation of covariance. This could be important for a large cross-correlation analysis or memory management during a large analysis, such as an LSST-like $3\times2$ analysis. 
This method provides a systematic means of integrating the scale cut applied in one space into another. This will be of particular importance for analyzing forthcoming data with improved constraining power. 
Future works should test the performance of this method using more complicated simulations and real data. Here we list targets of future work:
\begin{itemize}
    \item Testing the performance of this method using real cosmic shear data from existing surveys (e.g., HSC). 
    \item Extending this method to cosmological analyses that include scale cuts and survey windows (e.g., galaxy clustering, galaxy-galaxy lensing, CMB lensing, and $3\times2$pt)
    \item Understanding the impact of window biases on the reconstruction of the power spectrum. Window biases lead to both additive and multiplicative biases in the power spectrum. 
\end{itemize}
The future work outlined here will get this very promising method ready for direction application to Stage-IV surveys. 



\section*{Acknowledgments}
This work was performed using the Vera cluster at the McWilliams Center for Cosmology at Carnegie Mellon University, operated by the Pittsburgh Supercomputing Center facility. The authors would like to thank Scott Dodelson, Yuuki Omori, Rupert Croft, and Michael Troxel for many implementation discussions of the method and feedback on this work and thank Axel Guinot for thorough insightful comments on the draft of this paper.  We thank Shirasaki Masato for making the HSC Y1 shape mock catalog public. 

We thank the maintainers of numpy \citep{harris2020array}, \textsc{Jax} \citep{jax2018github}, Matplotlib \citep{Hunter:2007}, and \textsc{Healpix} \citep{2005ApJ...622..759G} for their excellent open-source software.

AP was partially supported by a McWilliams Center for Cosmology seed grant (PI: Singh).  SS was supported by the McWilliams postdoctoral fellowship at CMU. XL, RM, and TZ were supported in part by a grant from the Simons Foundation (Simons
Investigator in Astrophysics, Award ID 620789) and in part by the Department of Energy grant DE-SC0010118.

\section*{Data Availability}
This work used the public HSC Y1 mock shape catalogue, which can be found at \url{https://hsc-release.mtk.nao.ac.jp/doc/index.php/s16a-shape-catalog-data-products-pdr2/}. No new data were generated in support of this work. The software used for the computations of Wigner-3$j$ and coupling matrices from the survey window and the scale cut is available at \url{https://github.com/sukhdeep2/SkyLens/tree/skylens}.




\bibliographystyle{mnras}
\bibliography{main} 




\appendix
\onecolumn
\section{Cross-correlation between two reconstructions}
\label{sec:appendix:A}
\begin{figure}
    \centering
    \includegraphics[width=0.89\textwidth]{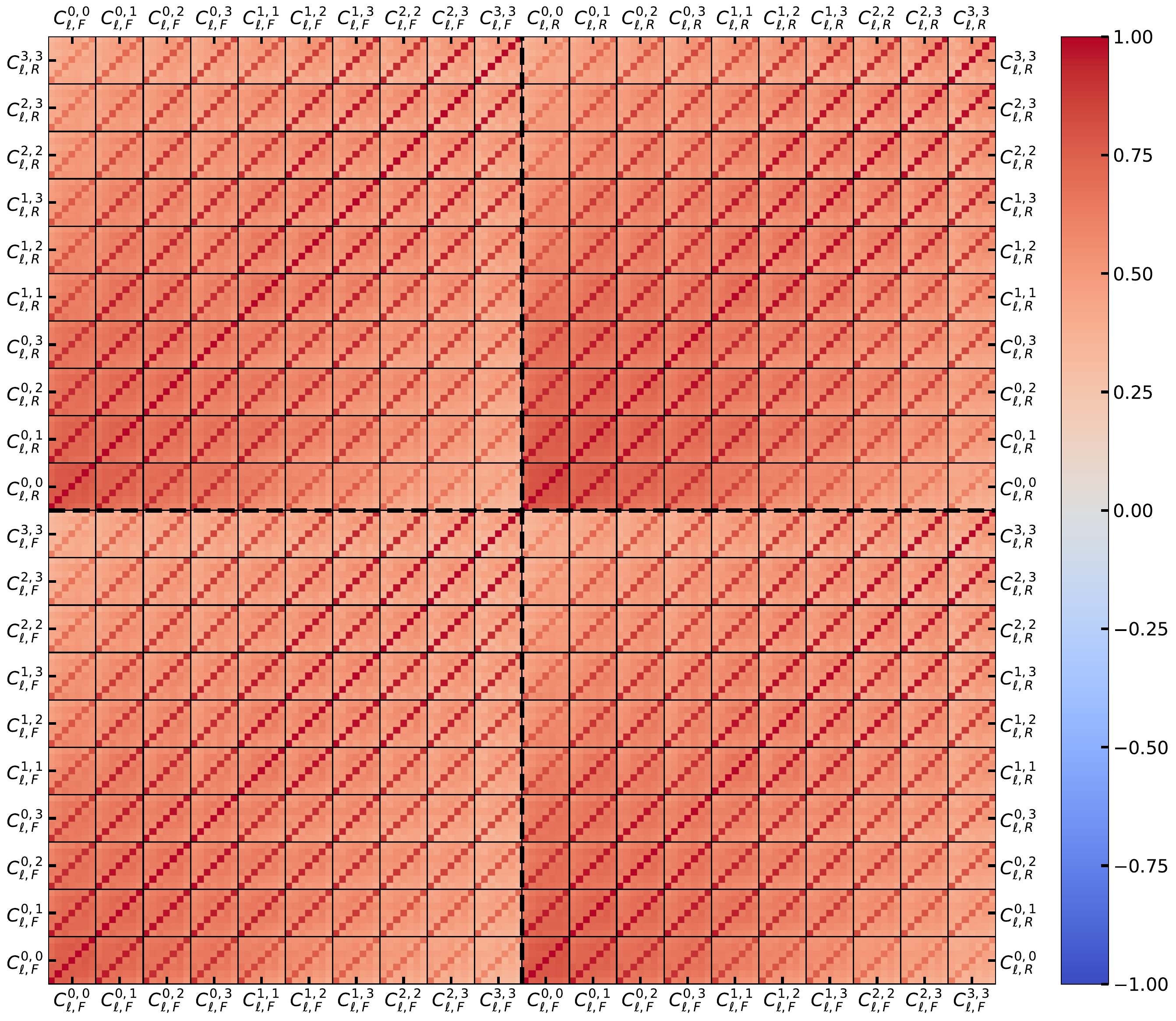}
    \caption{Cross-correlation (normalized covariance) between the two reconstructions where the lower-left and upper-right correlation blocks show the correlation matrix between two tomographic redshift bins within one reconstruction, and the lower-right and upper-left correlation blocks, which are identical, show the cross-correlation between the two reconstructions (using $x=\hat{C_\ell}$ in equation \ref{eq:result:corrcoef}). }
    \label{fig:crosscorr}
\end{figure}

In addition to showing the consistency between the two inferred values of $S_8$, we also show the consistency in the reconstruction of the power spectrum in Fig. \ref{fig:crosscorr}. Each block shows the value of $r(\hat{C}_\ell)$ defined in equation~\eqref{eq:result:corrcoef}. We find that all four blocks have homogeneous correlation coefficients, confirming the consistency between the two reconstructions of the power spectrum.

\newpage

\section{Scale Cuts and Survey Window}
\label{sec:appendix:B}
Here we show the derivation of the inverse spherical transform of two-point correlation functions measured in the presence of a survey window and a limited range of scales. This leads to the pseudo-$D_\ell$ described in equations~\eqref{eq:fell_ee} and~\eqref{eq:fell_bb}. Applying the inverse of the spherical transform in equation~\eqref{eq:xi_curved} gives
\begin{equation}
    F^{EE/BB}_\ell = \frac{1}{2}\int d\theta\:2\pi\sin\theta\left[{}_2 d^*_{\ell 2}(\theta) \xi_+(\theta)\xi^{(\gamma)}_W(\theta)\nu(\theta) \pm {}_{-2} d^*_{\ell 2}(\theta)\xi_-(\theta)\xi^{(\gamma)}_W(\theta)\nu(\theta)\right],
\end{equation}
where ${}_{s} d_{\ell, m}$ is the Wigner-d matrix, with superscript $*$ denoting complex conjugate and $s=m=2$ for spin-2 shear field, $\xi_W^{(\gamma)}(\theta)$ is the survey window correlation function defined in equation~\eqref{eq:method:xi_survey}, and $\nu(\theta)$ is the scale cut. Replacing the Wigner-d matrix with spin-weighted spherical harmonics (see discussions in \citealt{1999IJMPD...8...61N} and the Appendix in \citealt{2021MNRAS.508.1632S}), we get

\begin{align}
    F^{EE/BB}_\ell &= \frac{1}{2}\sqrt{\frac{4\pi}{2\ell+1}}\int d\Omega \left[{}_{2}Y^*_{\ell -2}(\theta, 0) \xi_+(\theta) \xi^{(\gamma)}_W(\theta)\nu(\theta) \pm {}_{2}Y^*_{\ell 2}(\theta, 0) \xi_-(\theta) \xi^{(\gamma)}_W(\theta)\nu(\theta)\right] \\
    \nonumber&=\frac{1}{2}\sqrt{\frac{4\pi}{2\ell+1}}\int d\Omega \sum_{\ell' \ell''}\sqrt{\frac{2\ell'+1}{4\pi}}\sqrt{\frac{2\ell''+1}{4\pi}}C^{(\nu)}_{\ell''}\left\{(D^{EE}_{\ell'} + D^{BB}_{\ell'}){}_{-2} Y_{\ell 2}(\theta) {}_{2} Y_{\ell' -2}(\theta) {}_{0} Y_{\ell'' 0}(\theta)\right.\\
    &\left.\quad\quad  \pm \:(D^{EE}_{\ell'} - D^{BB}_{\ell'}){}_{-2} Y_{\ell -2}(\theta) {}_{2} Y_{\ell' 2}(\theta) {}_{0} Y_{\ell'' 0}(\theta)\right\},
\end{align}
where $C^{(\nu)}_{\ell''}$ is the angular power spectrum of the scale cut $\nu(\theta)$. We used the identity ${}_{s}Y_{\ell m}^* = (-1)^{s + m} {}_{-s}Y_{\ell -m}$ and the fact that the inverse spherical transform of $\xi_\pm(\theta) \xi^{(\gamma)}_W(\theta)$ in the integrand is the pseudo-$D_\ell$ as shown in Appendix E of \cite{2021MNRAS.508.1632S}. Integrating the product of three spin-weighted spherical harmonics, we get

\begin{align}
    F^{EE/BB}_\ell &=\nonumber\frac{1}{2}\sqrt{\frac{4\pi}{2\ell+1}}\sum_{\ell' \ell''}\sqrt{\frac{2\ell'+1}{4\pi}}\sqrt{\frac{2\ell''+1}{4\pi}}C^{(\nu)}_{\ell''}\\
    &\nonumber\quad\quad\times\left\{D^{EE/BB}_{\ell'}\sqrt{\frac{(2\ell+1)(2\ell'+1)(2\ell''+1)}{4\pi}}\left[\begin{pmatrix}
        \ell & \ell' & \ell'' \\
        2 & -2 & 0
    \end{pmatrix}\begin{pmatrix}
        \ell & \ell' & \ell'' \\
        2 & -2 & 0
    \end{pmatrix} + \begin{pmatrix}
        \ell & \ell' & \ell'' \\
        2 & -2 & 0
    \end{pmatrix}\begin{pmatrix}
        \ell & \ell' & \ell'' \\
        -2 & 2 & 0
    \end{pmatrix}\right]\right.\\
    &\quad\quad\left.+\quad D^{BB/EE}_{\ell'}\sqrt{\frac{(2\ell+1)(2\ell'+1)(2\ell''+1)}{4\pi}}\left[\begin{pmatrix}
        \ell & \ell' & \ell'' \\
        2 & -2 & 0
    \end{pmatrix}\begin{pmatrix}
        \ell & \ell' & \ell'' \\
        2 & -2 & 0
    \end{pmatrix} - \begin{pmatrix}
        \ell & \ell' & \ell'' \\
        2 & -2 & 0
    \end{pmatrix}\begin{pmatrix}
        \ell & \ell' & \ell'' \\
        -2 & 2 & 0
    \end{pmatrix}\right]\right\},
\end{align}
where we used
\begin{equation}
    \int d\Omega~ {}_{s_1}Y_{\ell m_1}(\theta) {}_{s_2}Y_{\ell' m_2}(\theta) {}_{s_3}Y_{\ell'' m_3}(\theta) = \sqrt{\frac{(2\ell+1)(2\ell'+1)(2\ell''+1)}{4\pi}}\begin{pmatrix}
        \ell & \ell' & \ell'' \\
        -s_1 & -s_2 & -s_3
    \end{pmatrix}\begin{pmatrix}
        \ell & \ell' & \ell'' \\
        m_1 & m_2 & m_3
    \end{pmatrix}.
\end{equation}
Using the phase property of the Wigner-$3j$ symbols
\begin{equation}
    \begin{pmatrix}
        \ell & \ell' & \ell'' \\
        -s_1 & -s_2 & -s_3
    \end{pmatrix} = (-1)^{(\ell+\ell'+\ell'')}\begin{pmatrix}
        \ell & \ell' & \ell'' \\
        s_1 & s_2 & s_3
    \end{pmatrix}
\end{equation}
we derive the cosmic shear pseudo-$D_\ell$ as

\begin{align}
    \nonumber F^{EE/BB}_\ell &=\sum_{\ell'}\frac{2\ell'+1}{4\pi}\sum_{\ell''}(2\ell''+1)C^{(\nu)}_{\ell''}\left(\frac{1+(-1)^{\ell+\ell'+\ell''}}{2}\right)\begin{pmatrix}
        \ell & \ell' & \ell'' \\
        2 & -2 & 0
    \end{pmatrix}^2 \left(D^{EE/BB}_{\ell'} + D^{BB/EE}_{\ell'}\right)\\
    &= \sum_{\ell'}M^+_{\ell\ell'}D^{EE/BB}_{\ell'} + M^-_{\ell\ell'}D^{BB/EE}_{\ell'},
\end{align}
where the spin-2 coupling matrices are defined in equation~\eqref{eq:coup} with the power spectrum of survey window $C^{(W^\gamma)}_{\ell''}$ replaced with the power spectrum of the scale cut $C^{(\nu)}_{\ell''}$. If the full range of scales in $\ell$ and $\theta$ are used, the last equation is equivalent to showing that the power spectrum and two-point correlation function estimators are identical. 
%



\bsp	
\label{lastpage}
\end{document}